\documentclass{aastex}
\usepackage{spr-astr-addons}
\usepackage{url}

\RequirePackage{color}

\begin{document}

\title{\textit{XMM-Newton} and \textit{Swift} Observations of the Seyfert 1 AGN NGC~5940}
\slugcomment{Not to appear in Nonlearned J., 45.}
\shorttitle{XMM-Newton/Swift Observations of NGC~5940}
\shortauthors{Adegoke}

\author{Oluwashina K. Adegoke\altaffilmark{1}}

\altaffiltext{1}{Department of Mathematical \& Physical Sciences, Afe Babalola University, P.M.B 5454, Ado-Ekiti, Nigeria.}

\begin{abstract}
Probing the physics of the accretion flow around active galactic nuclei (AGN) is crucial to understanding their emission mechanisms as well as being able to constrain the geometrical and variability properties of the different regions around them. The soft X-ray excess -- usually observed below $\sim2\,\mathrm{keV}$ in excess of the dominant X-ray powerlaw continuum -- is one prominent feature that is commonly seen in type 1 Seyfert AGN and therefore readily provides a useful diagnostic of the accretion flow mechanism around these systems. \rm{NGC~5940} is a Seyfert 1 AGN which reveals strong, prominent soft X-ray excess below $\sim2\,\mathrm{keV}$ as seen in both its \textit{XMM-Newton} and \textit{Swift} observations. Model fit to the data revealed that this feature could be equally well explained by the ionised partial covering, the thermal Comptonisation and the blurred reflection models. Although the other models cannot be decisively ruled out with the data at hand, the lack of significant broad iron $K_{\alpha}$ as well as any significant emission/absorption line features in the reflection grating spectrometer (RGS) data tend to favour the thermal Comptonisation origin for the soft X-ray excess in \rm{NGC~5940}.
\end{abstract}



\section{Introduction}
Active galactic nuclei (AGN) have been known to be luminous in all bands of the electromagnetic spectrum. The optical/UV emission of AGN is believed to be produced in an optically thick standard accretion disc \citep{1973A&A....24..337S, 1973blho.conf..343N} due to viscous heating as material in the disc spiral down towards the central blackhole. The dominant X-ray powerlaw emission -- which is a significant fraction of an AGN's bolometric luminosity -- is theorised to be produced from the Compton-upscattering of seed disc optical/UV photons in a compact (size $\sim10-20\,R_{g}$, where $R_{g}$ is the gravitational radius), hot (having electron temperature $KT_{e}\sim100\,\mathrm{keV}$), optically thin (with optical depth $\tau<1$) electron plasma, called the corona \citep{1993ApJ...413..507H, 2019ApJ...870L..13A}. Other complex features imprinted on the X-ray spectra of AGN include the soft X-ray excess emission below $\sim2\,\mathrm{keV}$, iron $K_{\alpha}$ emission complex at $\sim6.4\,\mathrm{keV}$ and a Compton reflection ``hump'' which peaks normally at $\sim20\,\mathrm{keV}$.

 The soft X-ray excess is a smooth emission component in the $0.3-2.0\,\mathrm{keV}$ energy range that is commonly observed in Seyfert 1 AGN. It was first reported over three decades ago and was thought at the time to be the Wien's tail of the multi-temperature disc blackbody emission observed in the UV band \citep[see e.g.,][]{1985ApJ...297..633S, 1985MNRAS.217..105A, 1986MNRAS.218..685P, 1993A&A...274..105W}. However, it was soon shown that the predicted temperature ($\sim0.1-0.2\,\mathrm{keV}$) is much higher than expected for a standard accretion disc around a supermassive blackhole. It was equally found that the temperature is fairly constant and not related to the luminosity and mass of the central blackhole as against expectations that disc temperatures should be correlated with blackhole masses \citep[see e.g.,][]{2004MNRAS.349L...7G, 2005A&A...432...15P, 2009MNRAS.394..443M}. \citet{2004MNRAS.349L...7G} proposed that the soft excess could be an artefact of smeared absorption from partially ionised material, the model however requires unphysically large smearing velocity ($\sim0.5c$, where $c$ is the speed of light) for disc wind which has been shown to be difficult to generate from hydrodynamic simulations \citep{2007MNRAS.381.1413S}. Thus, the origin of the soft excess is still being debated even today. Most recently, two competing models have tried to explain the origin of the soft X-ray excess -- the thermal Comptonisation and the relativistically blurred reflection models. The thermal Comptonisation model posits that the soft X-ray excess results from the inverse-Compton scattering of disc optical/UV photons in an optically thick ($\tau \sim10$), warm ($KT_{e}\sim0.1-1\mathrm{keV}$) corona \citep[e.g.,][]{1998MNRAS.301..179M, 2012MNRAS.420.1848D, 2017MNRAS.466.3951A}, different from the hot ($\sim100\,\mathrm{keV}$) corona responsible for the dominant X-ray powerlaw. The relativistic reflection model on the other hand explains the soft X-ray excess to be the result of blurred reflection from partially ionised disc material in the vicinity of the blackhole due to its strong gravity and the high velocities of the accreting material \citep[e.g.,][]{2006MNRAS.365.1067C, 2014ApJ...782...76G, 2014MNRAS.444L.100D}. While both models tend to fit the data equally well in many cases, testing these models for a large number of sources and with data covering a wide range of the X-ray and other bands will be crucial to narrowing down on the actual origin of the soft X-ray excess as well as its possible dependence on the accretion flow properties of individual systems \citep[see e.g.,][]{2012MNRAS.425..907J, 2018A&A...611A..59P, 2020MNRAS.491..532G}. For such a large sample, a systematic search for the presence or absence of other features such as emission/absorption lines will help in identifying the preferred models. It is equally plausible that both thermal Comptonisation and X-ray reflection contribute simultaneously to the overall broadband X-ray spectra in AGN such that while the Compton reflection hump is due to relativistic reflection, thermal Comptonisation produces the soft X-ray excess, as was recently reported for the Seyfert 1 AGN \rm{Ton~S180} \citep{2020MNRAS.497.2352M}. 

\rm{NGC~5940} is a nearby Seyfert 1 AGN at a redshift of $0.034$ and luminosity distance of $147\,\mathrm{MPc}$ \citep{2019ApJ...880...68R}. Its mass has been estimated to be $1.085^{+0.182}_{-0.167}\times10^{7}\,M_{\odot}$ in the reverberation mapping database\footnote{{http://www.astro.gsu.edu/AGNmass/}} 
of \citet{2015PASP..127...67B}. This is based on the 2011 Lick Observatory $\rm{H}\beta$ and $\rm{Fe}II$ lag measurements reported in \citet{2013ApJ...769..128B, 2015ApJS..217...26B}. Despite its relative proximity, its spectral properties have not been probed in detail until now, especially in the X-ray energy band.

This paper presents a detailed analysis of the \textit{XMM-Newton} X-ray data of \rm{NGC~5940} -- supplemented with data from the \textit{Swift} X-ray Telescope. In Section 2, we describe the observation and data reduction procedure. In Section 3, we detail the data analysis approach used. We discuss the data analysis result in Section 4, and in Section 5, we draw conclusions and summarise our results.

\section{Observations and Data Reduction}
\rm{NGC~5940} was observed by \textit{XMM-Newton} \citep{2001A&A...365L...1J} on the 23rd of February, 2012 for a duration of $\sim33\,\mathrm{ks}$ (ObsID: 0670040601). All cameras onboard the telescope were used during the observation. The European Photon Imaging Cameras (EPIC) \citep{2001A&A...365L..18S, 2001A&A...365L..27T}  were operated in the full frame mode. While EPIC-pn was operated with thin filter, both MOS1 and MOS2 were operated with medium filters. The optical monitor (OM) \citep{2001A&A...365L..36M} was operated in the image mode using the UVM2 and UVW1 filters.

The data was processed using the Science Analysis System ({\tt sas v.17.0.0}) with updated Current Calibration Files (CCF). The sas task EVSELECT was employed to extract the event list. The data were subsequently screened and filtered for intervals of high particle background after which a good time interval (GTI) file was created. To check for possible pile-up in the data, a filtered event list was extracted from the raw event list -- from a radius of 40 arcseconds centred on the source coordinates -- using the GTI file after which the EPATPLOT command was applied. The check revealed the presence of significant pile-up in the pn data. To correct for the pile-up, X-ray loading was first corrected for, -- since X-ray loading normally accompanies piled-up observations -- after which the EVSELECT command was applied on the new event list using the GTI file to create a cleaned event file. Source photons were extracted from a region of radius 40 arcseconds centred on the source position while background was extracted from a circular source-free region of radius 80 arcseconds on the same CCD. The EVSELECT task was used to generate the X-ray source and background spectra.
  The SAS tasks ARFGEN and RMFGEN were subsequently used to compute the ancillary and photon redistribution matrices. Pile-up correction was implemented in the generation of the redistribution matrix which was calculated from the frequency and spectrum of the incoming photons. The resulting spectra were then grouped using the SPECGROUP command to have a minimum of 25 counts per bin to facilitate use of the $\chi^{2}$ minimisation technique.

The reduction of the reflection grating spectrometer (RGS) \citep{2001A&A...365L...7D} data followed standard procedure employing the SAS task RGSPROC. The accuracy of the source coordinates was checked from the source list and found to be in order. The data were subsequently checked against periods of high particle background by plotting the lightcurve of the pure background events. For improved signal-to-noise ratio, first order RGS1 and RGS2 spectra were combined using the task RGSCOMBINE and then grouped to have a minimum of 30 counts per bin for use with {\tt XSPEC} $\chi^{2}$ statistics.

\textit{Swift} \citep{2004ApJ...611.1005G} has observed \rm{NGC~5940} a couple of times since 2008. Although none of the snapshots were simultaneous with the 2012 \textit{XMM-Newton} observation of the source, four observations carried out between 2010 and 2011 were combined into one spectra of total exposure $\sim9.8\,\mathrm{ks}$ and used for spectral analysis. This is based on the assumption that the spectral properties of the source had not changed significantly between the time of the 2012 \textit{XMM-Newton} and the earlier 2010/2011 \textit{Swift} observations. The \textit{Swift} X-ray Telescope \citep[XRT;][]{2005SSRv..120..165B} data reduction was carried out with the aid of the online \textit{Swift}/XRT product generator hosted at the University of Leicester\footnote{{https://www.swift.ac.uk/index.php}} \citep{2009MNRAS.397.1177E}. 

\section{Spectral Analysis}
We used the EPIC-pn X-ray data for our spectral analysis due to its higher signal-to-noise ratio compared to EPIC-MOS. We carried out the spectral analysis on \rm{NGC~5940} using {\tt XSPEC v.12.11.1} \citep{1996ASPC..101...17A}. We used the $\chi^{2}$ statistics and for a single parameter of interest, errors were quoted at the 90\% confidence level -- corresponding to $\Delta\chi^{2}=2.706$ -- unless otherwise stated. In all the fitting, we have added contributions from galactic absorption column of $N_{H}=3.6\times10^{20}\,\mathrm{cm^{-2}}$ obtained from the recent HI21cm measurements \citep{2016A&A...594A.116H}, modelled with {\tt tbabs} \citep{2000ApJ...542..914W}.
\begin{figure}[!htb]
  \begin{center}
    \includegraphics[width=0.30\textwidth, angle=-90]{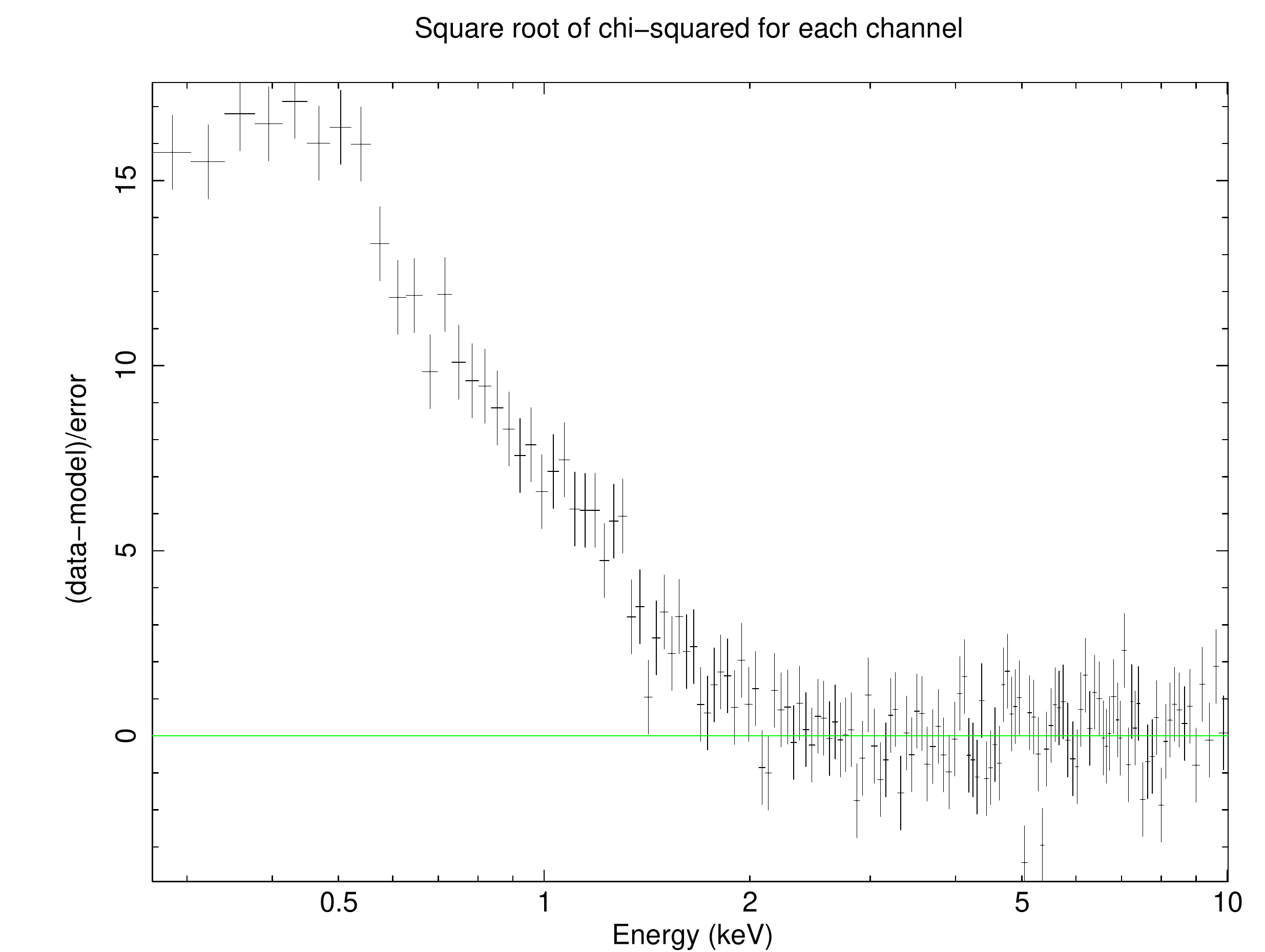}
    \caption[]{Residual plot revealing strong soft X-ray excess below $2\,\mathrm{keV}$ based on a powerlaw fit to the $2-10\,\mathrm{keV}$ EPIC-pn data of \rm{NGC~5940} extrapolated to $0.3\,\mathrm{keV}$.}
    \label{sxe}
  \end{center}
\end{figure}

We began by fitting the $2-10\,\mathrm{keV}$ band with a baseline model consisting of a simple powerlaw modified by galactic absorption (i.e. {\tt tbabs*powerlaw}). This provided a very good fit to the data with photon index $\Gamma=1.78$ and $\chi^{2}/dof$ of 91.7/93, where $dof$ represents the degrees of freedom. Inclusion of a gaussian line to model possible iron $K_{\alpha}$ features did not improve the fit, giving $\chi^{2}/dof=88.3/90$. 
When extrapolated down to $0.3\,\mathrm{keV}$, the data revealed strong soft X-ray emission in the source below $\sim2\,\mathrm{keV}$ in excess of the powerlaw as shown in the residual plot of Fig. \ref{sxe}, providing an unacceptable fit with $\chi^{2}/dof$ of 223.6/140. To test existing models on the origin of this soft excess emission, we fitted the entire $0.3-10\,\mathrm{keV}$ band with phenomenological and subsequently physically motivated models.  

\subsection{Phenomenological Model}
To model the soft X-ray excess, we added the multicolour disc blackbody model {\tt diskbb} \citep[see e.g.,][]{1984PASJ...36..741M, 1986ApJ...308..635M}  to the baseline model (i.e. {\tt tbabs*(diskbb+powerlaw)}). This provided an acceptable fit to the data with $\chi^{2}/dof=159.7/138$. The inner disc temperature $T_{in}\sim0.18\,\mathrm{keV}$ is high for an accretion disc around a $10^{7}\,M_{\odot}$ blackhole. The photon index $\Gamma$ is 1.89. Adding a gaussian line at $6.4\,\mathrm{keV}$ to the model slightly improved the fit with $\Delta\chi^{2}/\Delta dof=-12/-3$, giving $\chi^{2}/dof=147.4/135$. However, in the subsequent models, addition of a gaussian line at $\sim6.4\,\mathrm{keV}$ did not improve the fits in any way and was therefore excluded. 
\begin{figure*}
\centering
\includegraphics[width=.23\textwidth, height=0.23\textheight, angle=-90]{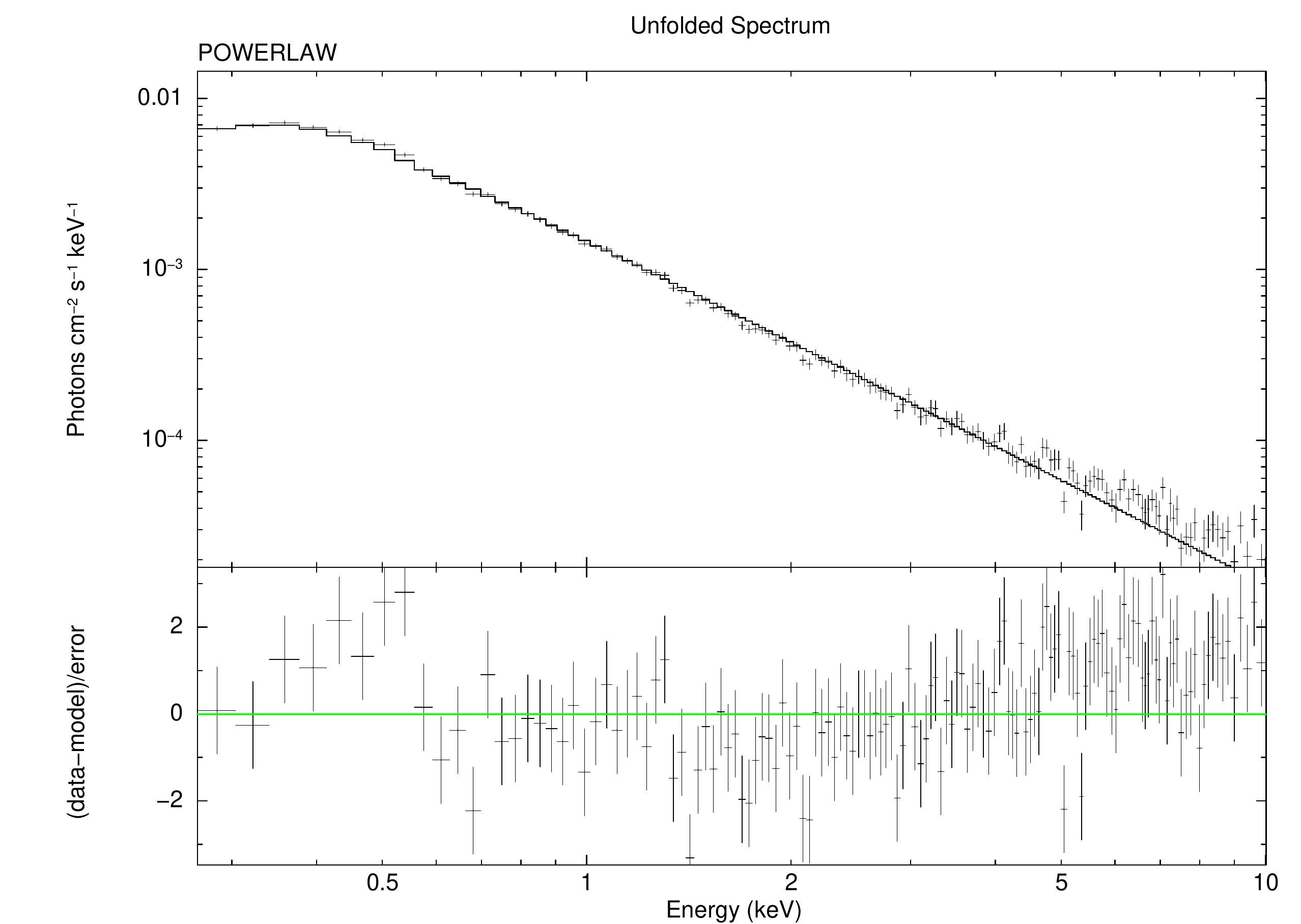}
\includegraphics[width=.23\textwidth, height=0.23\textheight, angle=-90]{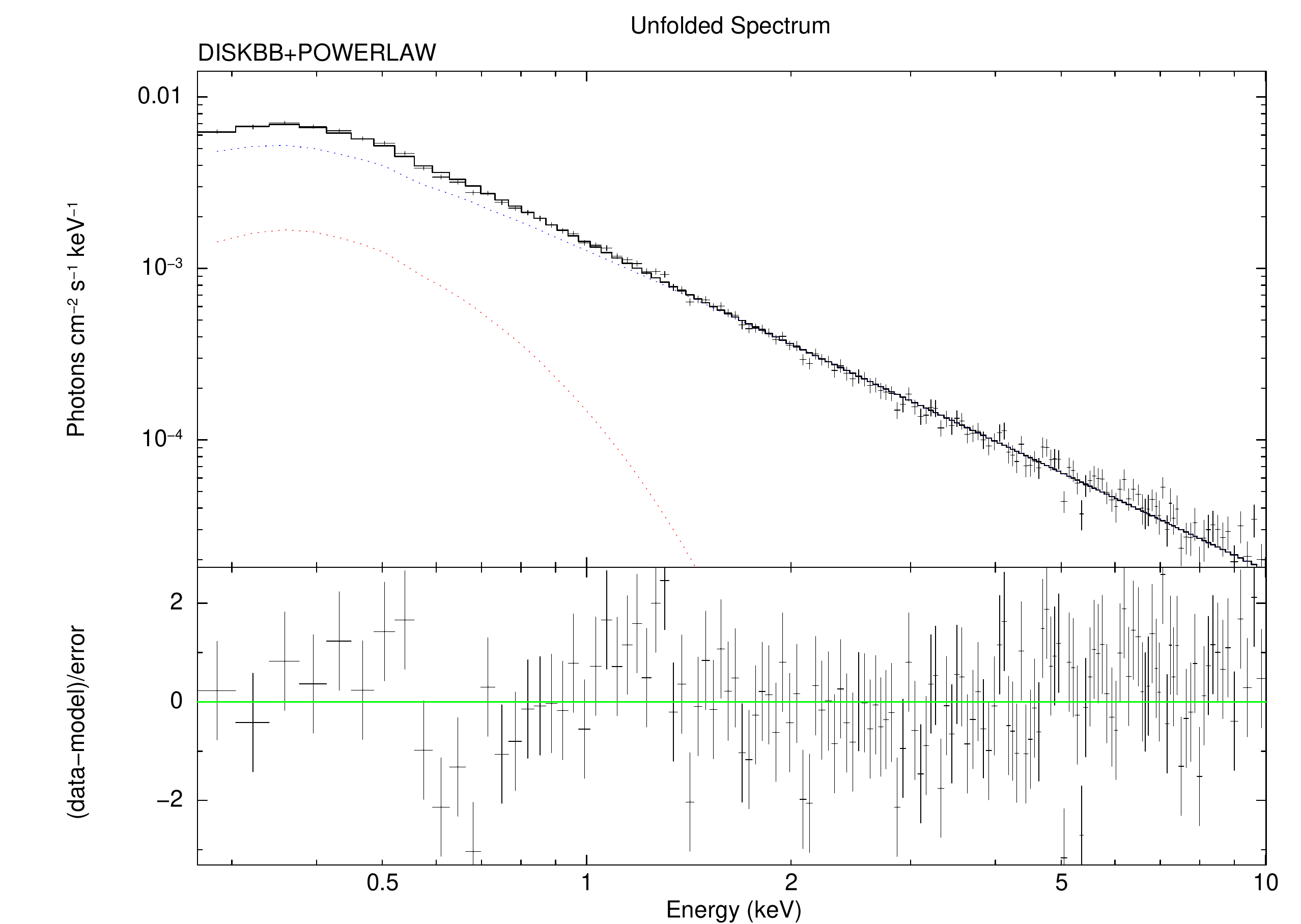}  
\includegraphics[width=.23\textwidth, height=0.23\textheight, angle=-90]{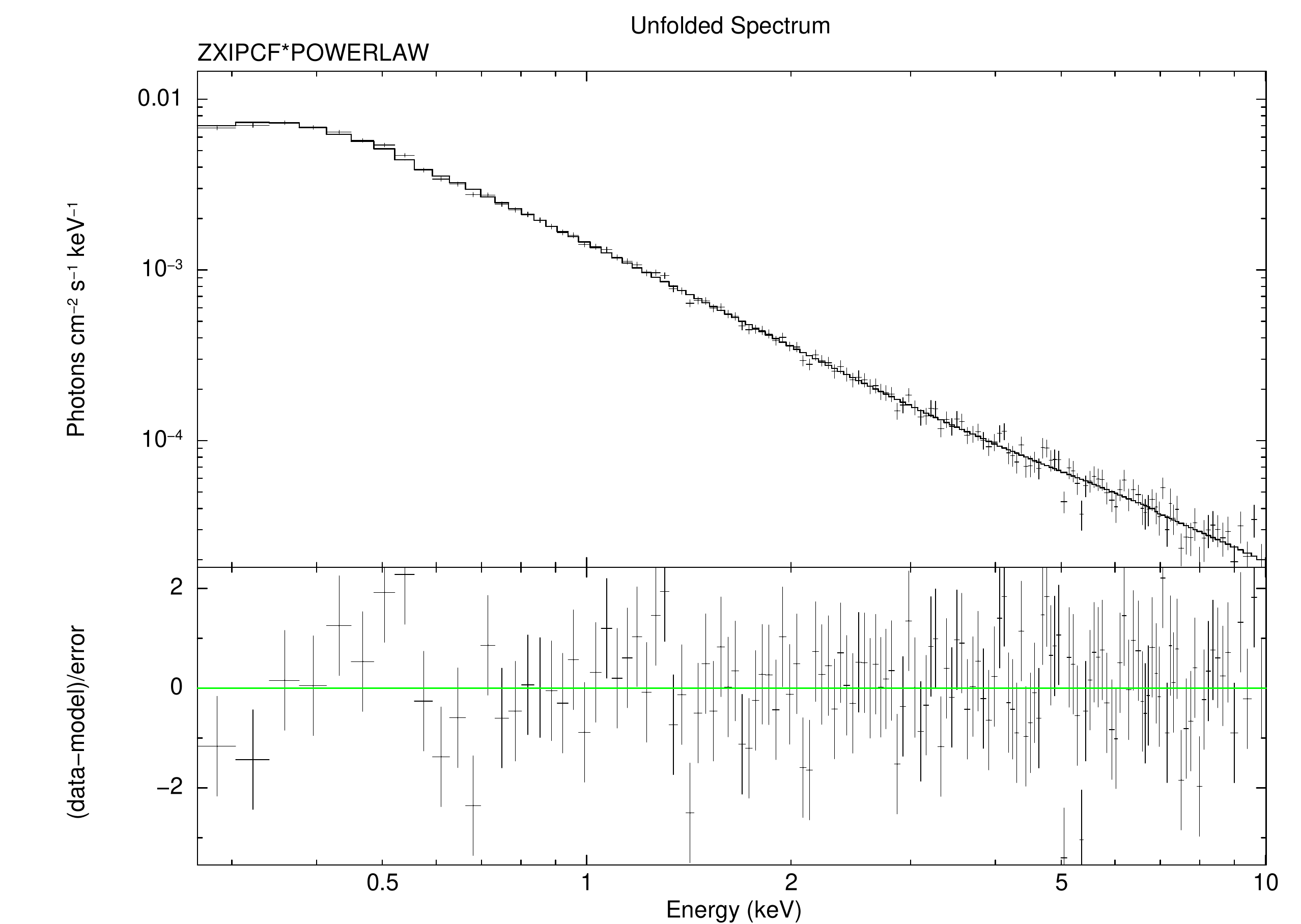} 
\includegraphics[width=.23\textwidth, height=0.23\textheight, angle=-90]{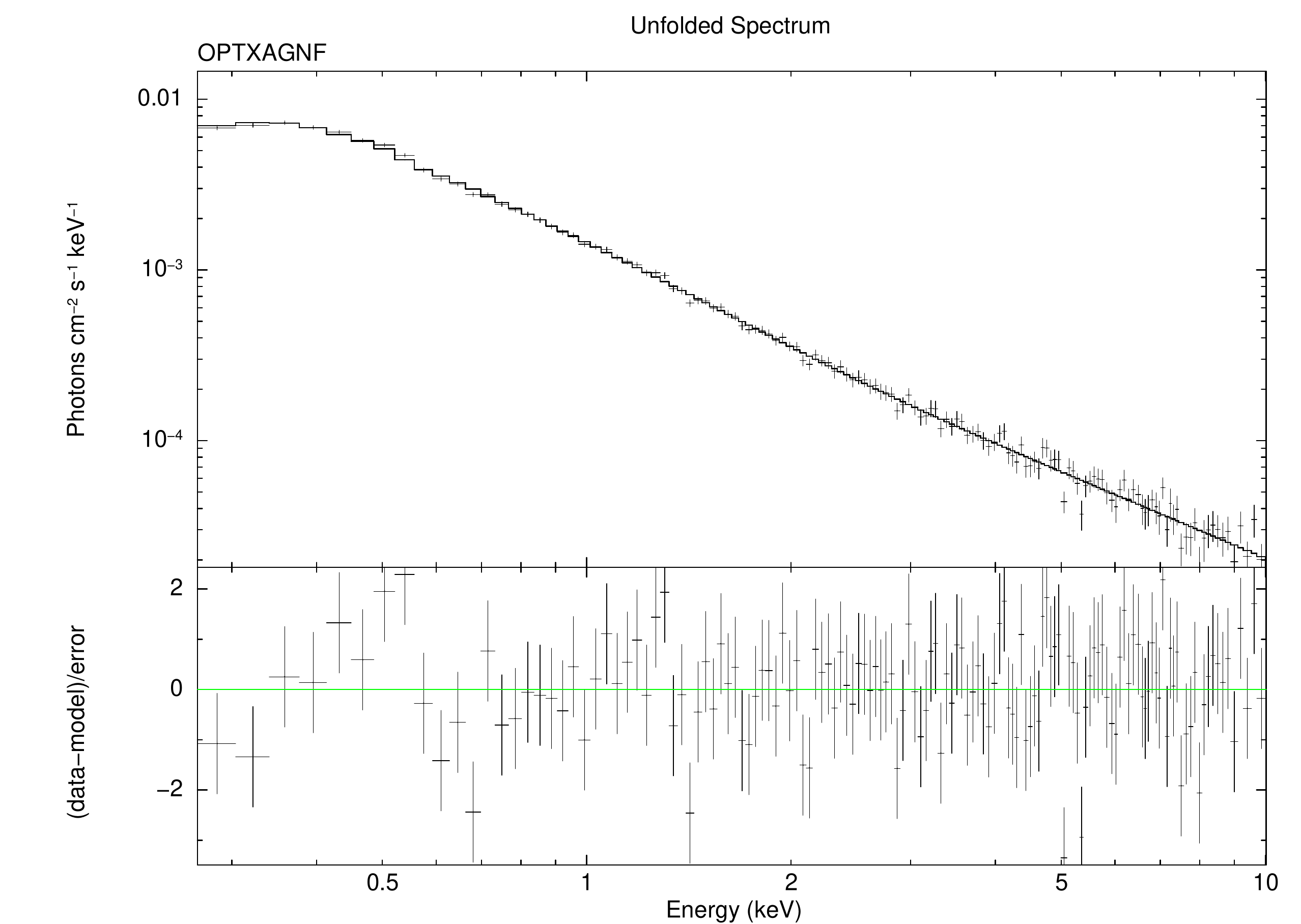}
\includegraphics[width=.23\textwidth, height=0.23\textheight, angle=-90]{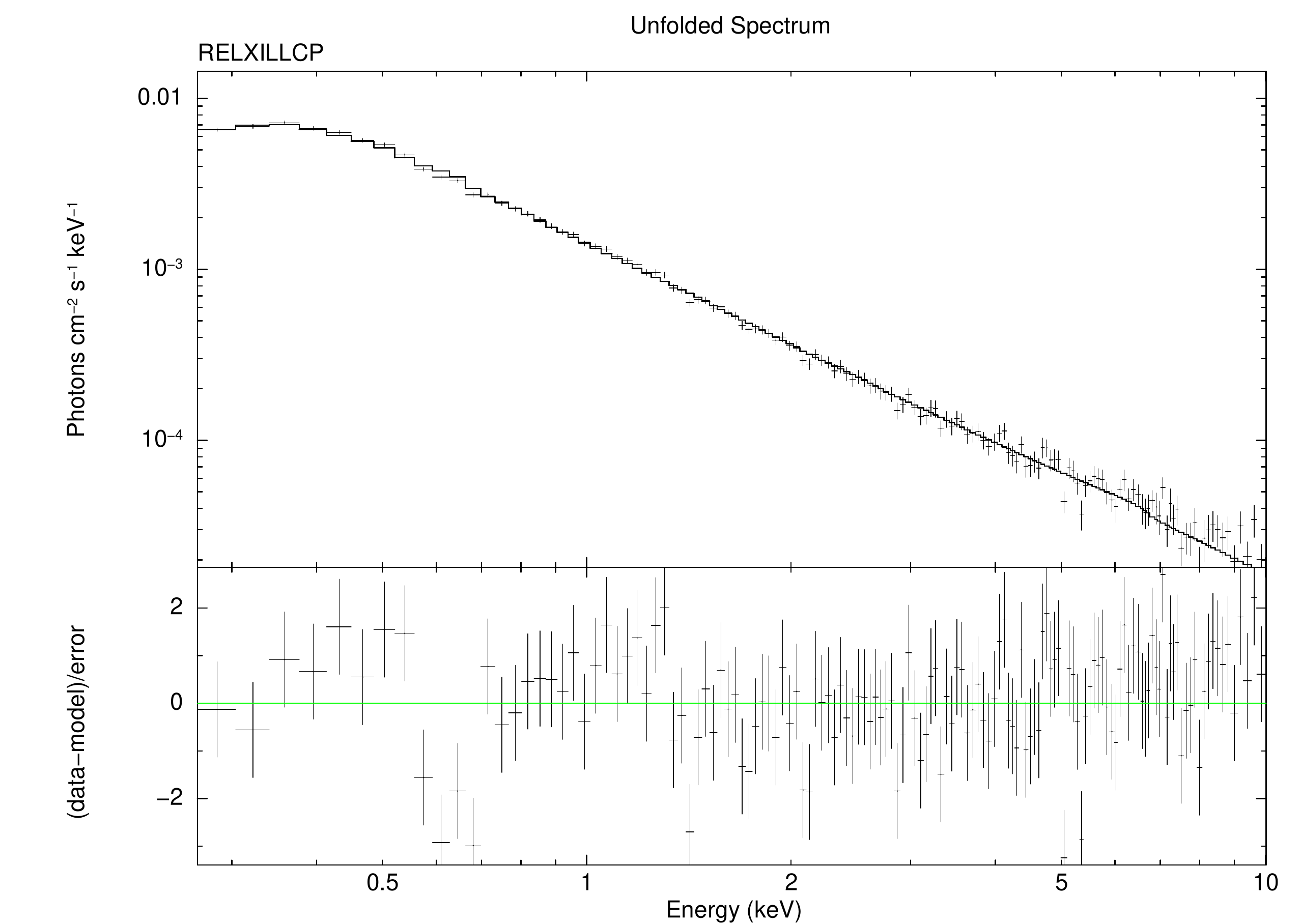} 
\includegraphics[width=.23\textwidth, height=0.23\textheight, angle=-90]{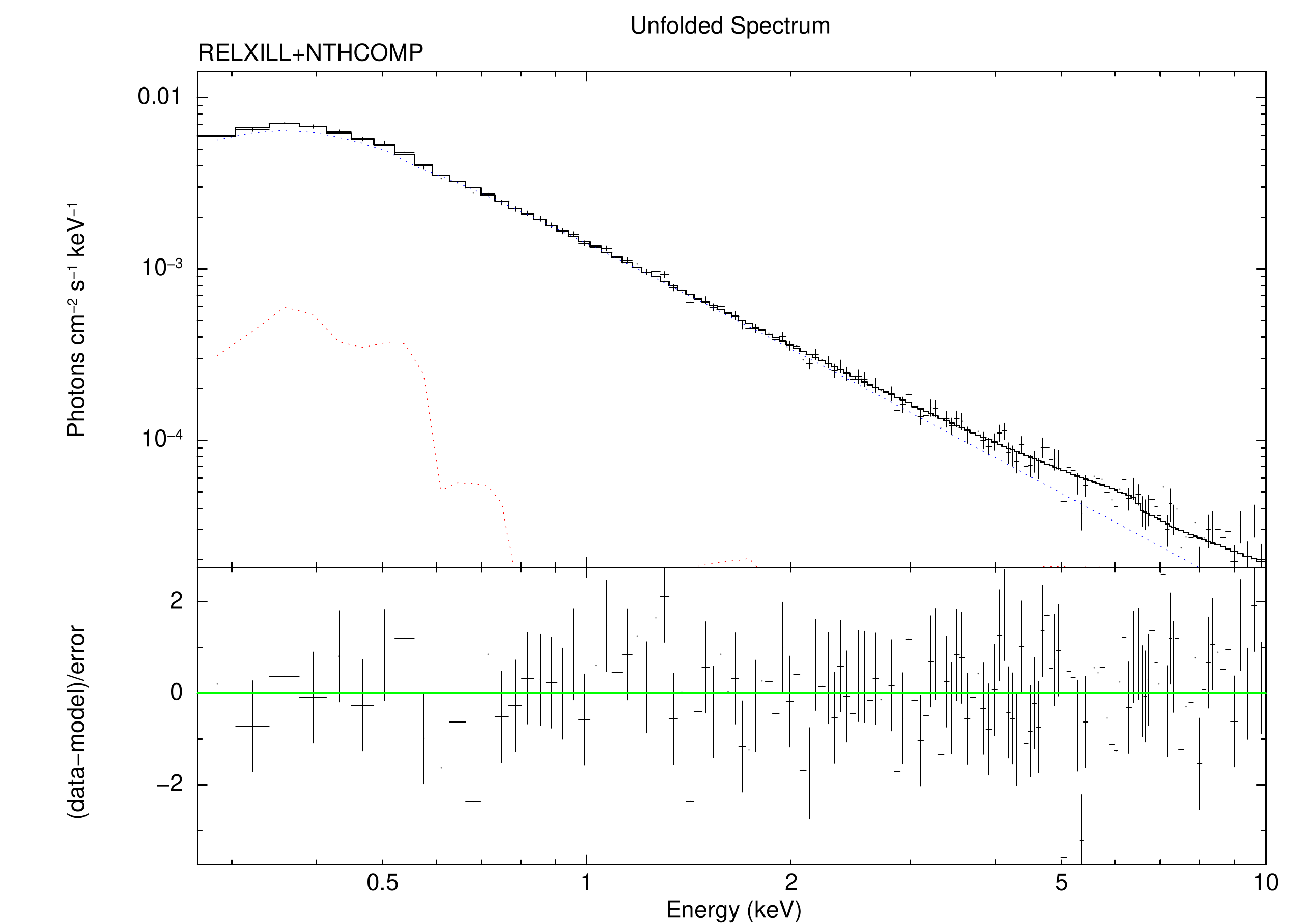}
\caption{Unfolded spectra of best fit for each of the models applied to the $0.3-10\,\mathrm{keV}$ \textit{XMM-Newton} EPIC-pn data of \rm{NGC~5940}. The upper panels in each plot show the model-data fit -- the black solid lines represent the composite model in each of the plots. The lower panels in each plot show the residuals.}
\label{spec_fit}
\end{figure*}
\subsection{Physical Models}
\subsubsection{Ionised partial covering: zxipcf}
The shape of the residual plot for the powerlaw fit to the $0.3-10\,\mathrm{keV}$ data seem to show features that may result from complex partial covering absorption along the line of sight (shown in Fig. \ref{spec_fit} -- top, leftmost plot). In this scenario, the intrinsic powerlaw continuum is obscured by an ionised absorber along the line of sight. The observed spectrum is then a combination of the obscured and the direct powerlaw components \citep[see e.g.,][]{2008MNRAS.385L.108R, 2011MNRAS.412..161G, 2019MNRAS.488.4831T}. We tested this possibility by multiplying the absorbed powerlaw with the ionised partial covering absorption model {\tt zxipcf} (i.e. {\tt tbabs*zxipcf*powerlaw}). This provided a very good fit to the data with $\chi^{2}/dof=135.9/137$, covering fraction $C_{f}=0.42$ and line of sight column density $N_{h}\sim1.6\times10^{23}\,\mathrm{cm^{-2}}$, the best-fit parameters are shown in Table \ref{table11}. 
\begin{table*}
\caption{The best-fit parameter values from fitting different models to the $0.3-10\,\mathrm{keV}$ \textit{XMM-Newton} EPIC-pn data of \rm{NGC~5940}.} 
\centering 
\begin{tabular}{l l} 
\hline\hline 
\textbf{Model/Parameter} \& \textbf{Best-fit values} \\ [0.5ex] 
\hline\hline 
Model & {\tt tbabs*(diskbb+powerlaw)}\\
$T_{\mathrm{in}}(\mathrm{keV})$ & $0.185\pm{0.02}$ \\  
Normalisation - diskbb & $51.28^{+24.41}_{-15.51}$ \\
Photon index $\Gamma$ & $1.89\pm0.03$ \\
Normalisation ($10^{-3}$) - powerlaw & $1.35\pm0.05$ \\
$\chi^2/dof$ & 159.7/138 \\ 
\hline
Model & {\tt tbabs*(zxipcf*powerlaw)}\\
Column density $N_{h}$ ($10^{22}$) - zxipcf & $15.74^{+24.49}_{-9.96}$ \\
Ionisation parameter log($\xi$) & $-0.20^{+1.41}_{-P}$ \\
Covering fraction $C_{Frac}$ & $0.42^{+0.15}_{-0.09}$ \\
Photon index $\Gamma$ & $2.09\pm0.02$ \\
Normalisation ($10^{-3}$) - powerlaw & $2.62^{+0.96}_{-0.34}$ \\
$\chi^2/dof$ & 135.9/137 \\ 
\hline
Model & {\tt tbabs*(optxagnf)}\\
Mass ($\times10^{7}\,M_{\odot}$) & $1.1$(f) \\
Luminosity distance ($\mathrm{MPc}$) & $147$(f) \\
log($\frac{L}{L_{Edd}}$) & $-1.26^{+0.60}_{-0.06}$ \\
spin a* & $0.998$(f) \\
Corona radius $R_{cor}$ & $18.12^{+P}_{-11.14}$ \\
Warm corona $kT_{SE}$ ($\mathrm{keV}$) & $0.40^{+0.14}_{-0.09}$ \\
Optical depth $\tau$ & $12.92^{+2.55}_{-2.05}$ \\
Photon index $\Gamma$ & $1.71^{+0.09}_{-0.12}$ \\
Powerlaw fraction $f_{pl}$ & $0.77^{+0.08}_{-0.21}$ \\
$\chi^2/dof$ & 135.6/136 \\ 
\hline
Model & {\tt tbabs*(relxillcp)}\\

spin a* & $0.998$(f) \\
Inclination & $30\deg$(f) \\
Photon index $\Gamma$ & $1.92\pm0.02$ \\
Ionisation parameter log($\xi$) & $3.00^{+0.10}_{-0.15}$ \\
Iron abundance $A_{Fe}$ & $1.0$(f) \\
Hot corona $kT_{e}$ ($\mathrm{keV}$) & $100$(f) \\
Reflection frac & $0.55^{+0.16}_{-0.13}$ \\
Normalisation ($10^{-5}$) & $1.89^{+0.11}_{-0.14}$ \\
$\chi^2/dof$ & 164/138 \\ 
\hline 
Model & {\tt tbabs*(relxill+nthcomp)}\\

spin a* & $0.991^{+P}_{-0.418}$ \\
Inclination & $30\deg$(f) \\
Photon index $\Gamma$- relxill & $2.12^{+0.01}_{-0.04}$ \\
Ionisation parameter log($\xi$) & $0.02^{+0.30}_{-P}$ \\
Iron abundance $A_{Fe}$ & $1.0$(f) \\
High energy cutoff $E_{cut}$ ($\mathrm{keV}$) & $300$(f) \\
Reflection frac $R_{frac}$ & $-1$(f) \\
Normalisation ($10^{-5}$) - relxill & $5.65^{+0.92}_{-1.57}$ \\
Photon index $\Gamma$ - nthcomp & $2.12$(t) \\
Hot corona $kT_{e}$ ($\mathrm{keV}$) & $100$(f) \\ 
Disk blackbody $kT_{bb}$ ($\mathrm{keV}$) & $0.1$(f) \\
Normalisation ($10^{-3}$) - nthcomp & $1.49\pm0.02$ \\
$\chi^2/dof$ & 134.1/138 \\ 
\hline 
\hline 

\end{tabular}
\label{table11} 

{Note: ``f'' implies that the value is frozen, ``t'' implies that the parameter value is tied to another parameter and ``P'' implies that the error value is pegged at the hard limit.}
\end{table*}

\subsubsection{Warm corona model: optxagnf}
The warm coronal model \citep{2012MNRAS.420.1848D} posits that the gravitational energy released at each annulus of the disc is radiated as thermal blackbody emission down to the radius of the corona $R_{cor}$. The gravitational potential energy within this radius is assumed to no longer be completely thermalised, giving rise to a warm ($kT_{SE}\sim0.2\,\mathrm{keV}$), optically thick ($\tau>1$) plasma responsible for the soft X-ray excess emission and a hot ($kT_{e}\sim100\,\mathrm{keV}$), optically thin ($\tau<1$) corona responsible for the dominant X-ray powerlaw continuum above $\sim2\,\mathrm{keV}$. Important parameters of the model include the blackhole mass $M_{BH}$, the dimensionless spin a*,  the electron temperature $kT_{SE}$ for the soft Comptonisation component and its optical depth $\tau$, the photon index of the powerlaw X-ray continuum $\Gamma$, the Eddington ratio in log scale log($L/L_{Edd}$) and the luminosity distance $D_{L}$ to the source. In fitting with this model, we fixed the blackhole mass at $\sim1.1\times10^{7}\,M_{\odot}$ \citep{2019ApJ...880...68R}, the disc outer radius at its default value and  $D_{L}$ at $147\,\mathrm{MPc}$. Following \citet{2005MNRAS.361.1197C}, we fixed and froze the spin parameter to the maximum allowed value of a*$=0.998$ as it was found to be unconstrained when freed. This model resulted in an excellent fit with $\chi^{2}/dof=135.6/136$. The best-fit values of the Eddington ratio, $R_{cor}$, $kT_{SE}$, optical depth and the powerlaw photon index are $0.05$, $18.08\,R_{g}$, $0.40\,\mathrm{keV}$, $12.92$ and $1.71$ respectively. All the fit parameters are shown in Table \ref{table11}.
\subsubsection{Blurred reflection: relxill}
To a good extent, reflection models have been successful in modelling some of the prominent features imprinted on the X-ray spectrum of AGN including the soft X-ray excess, the hard X-ray excess as well as line profiles like the iron $K_{\alpha}$ emission feature seen at $\sim6.4\,\mathrm{keV}$ \citep[see e.g.,][]{2004MNRAS.353.1071F, 2013MNRAS.428.2901W, 2019MNRAS.489.3436J}. To test whether blurred relativistic reflection in the immediate vicinity of the AGN can explain the origin of the prominent soft X-ray excess emission, we used the {\tt relxill} model. This model combines the capabilities of the reflection code {\tt xillver} \citep{2014ApJ...782...76G} with the relativistic ray-tracing code {\tt relline} \citep{2014MNRAS.444L.100D}. It models the irradiation of the accretion disc by a broken powerlaw emissivity of the form $\epsilon\propto r^{-q_{1}}$ between $r_{in}$ and $r_{br}$ and $\epsilon\propto r^{-q_{2}}$ between $r_{br}$ and $r_{out}$; where $q_{1}$ and $q_{2}$ are the inner and outer emissivity indices, $r$ is the radius of the accretion disc while $r_{in}$, $r_{br}$ and $r_{out}$ are the inner, break and outer radii of the accretion disc respectively. More information about the model and its different flavours are available on its documentation webpage\footnote{\url{http://www.sternwarte.uni-erlangen.de/~dauser/research/relxill/}}. We started by applying the {\tt relxillcp} flavour of the model which is more physically consistent as the reflection fraction in this case is computed using the primary continuum implemented with the {\tt nthcomp}  Comptonisation model \citep{1996MNRAS.283..193Z, 1999MNRAS.309..561Z}.

During fitting, the indices $q_{1}$ and $q_{2}$ were kept at their default values of $3$. Also, $r_{br}$, inclination, $r_{in}$ and $r_{out}$ were left frozen at their default values. Iron abundance relative to solar abundance $A_{Fe}$ and the hot corona temperature were fixed at $1$ and $100\,\mathrm{keV}$ respectively. The spin parameter a* was fixed at its maximum value of 0.998. The model provided an acceptable fit to the data with $\chi^2/dof=164/138$ and a steeper photon index $\Gamma=1.92$. Allowing the inner emissivity index $q_{1}$, break radius $r_{br}$, spin, inclination and the iron abundance $A_{Fe}$ parameters to be free improved the fit significantly, with $\Delta\chi^{2}/\Delta dof=-20/-4$ giving $\chi^2/dof=144/133$. This configuration additionally gave $r_{br}\sim4\,R_{g}$, $q_{1}\sim9.8$, a*$=0.899$, $\Gamma=1.96$, inclination $\sim35\deg$ and $A_{Fe}\sim4$. As evident, the inner emissivity index is rather extreme in this case which may indicate strong inner disc reflection due to strong general relativistic effects in the immediate vicinity of the blackhole. The modelled iron abundance is also considerably super-solar.

We afterwards replaced the {\tt relxillcp} with the {\tt relxill+nthcomp} model where the illuminating powerlaw is modelled with {\tt nthcomp} and the blurred reflection only with {\tt relxill}. For this, the reflection fraction parameter $R_{frac}$ was set to $-1$. Most parameters of the {\tt relxill} and {\tt relxillcp} models are the same with the exception of the coronal electron temperature $kT_{e}$ in {\tt relxillcp} which is replaced by the high energy cutoff $E_{cut}$ in {\tt relxill}. The photon index of {\tt nthcomp} was tied to that of the {\tt relxill} model. This provided an excellent fit to the data with $\Delta \chi^{2}=-10$ compared to the previous model for four additional degrees of freedom, giving $\chi^2/dof=134.1/137$ and a slightly steeper photon index value $\Gamma=2.12$. Table \ref{table11} contains all the fitted parameter values of the model.      

\subsection{RGS spectral analysis}
\begin{figure}[!htb]
  \begin{center}
    \includegraphics[width=0.35\textwidth, angle=-90]{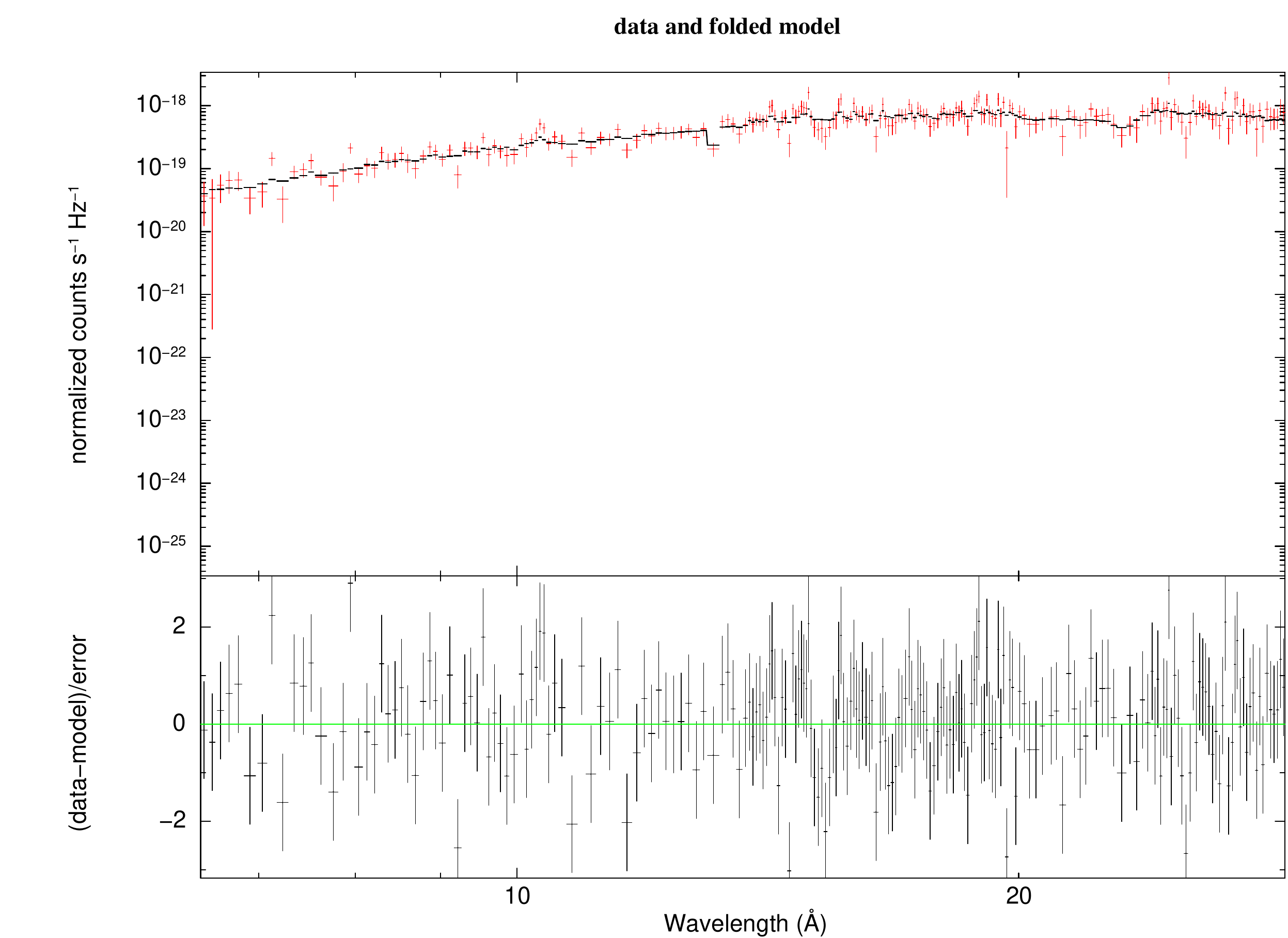}
    \caption[]{The upper panel shows a powerlaw fit (in black) to the first order RGS1+2 spectrum (in red) of \rm{NGC~5940}. The lower panel shows the residual plot.}
    \label{rgs}
  \end{center}
\end{figure}

We analysed the RGS spectra of \rm{NGC~5940} to check for possible prominent emission and absorption line features in the $0.3-2\,\mathrm{keV}$ energy range associated with the soft X-ray excess. This is because the RGS has a much higher energy resolution compared to both EPIC-pn and MOS and as such these reflection features should be evident in the RGS spectra of the source if blurred relativistic reflection is predominantly what produces the soft X-ray excess. We modelled the combined first order RGS spectra with a simple powerlaw modified by galactic absorption. This provided a very good fit with $\chi^2/dof=216.9/221$ and $\Gamma=2.12$. Figure \ref{rgs} shows the powerlaw modelled RGS (1+2) stacked spectra of \rm{NGC~5940}, with the residual shown in the lower panel of the plot. This spectral analysis result suggested that the soft X-ray spectra in \rm{NGC~5940} is largely featureless, consistent with a simple powerlaw. Following \citet{2020MNRAS.497.2352M}, we systematically scanned blindly the $6.5-29\,\mathrm{\textup{\AA}}$ wavelength band of the RGS spectra for absorption and emission line features. To do this, we included a Gaussian line profile in the absorbed powerlaw model, allowing for both positive and negative normalisations. The search did not result in the detection of any prominent features that is significant.

\subsection{\textit{Swift} XRT spectral analysis}
\begin{figure}[!htb]
  \begin{center}
    \includegraphics[width=0.35\textwidth, angle=-90]{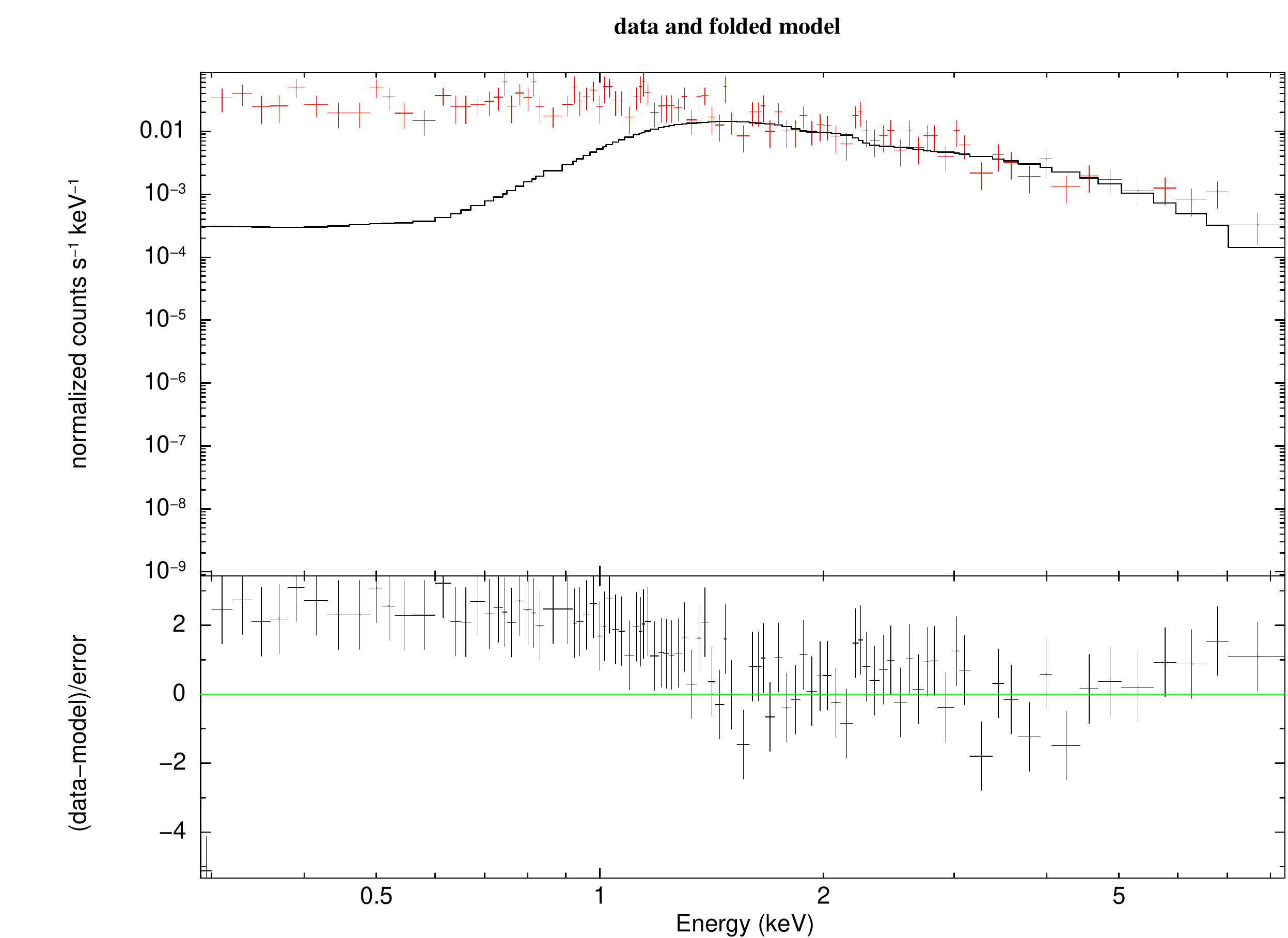}
    \caption[]{The upper panel shows a powerlaw fit (in black) to the $1.5-7.0\,\mathrm{keV}$ \textit{Swift} XRT spectrum (in red) of \rm{NGC~5940} extrapolated down to $0.3\,\mathrm{keV}$. The lower panel shows the residual plot.}
    \label{swift}
  \end{center}
\end{figure}

We combined \textit{Swift} XRT spectra from four observations (ObsIDs: 00037386002, 00037386003, 00037386004, 00037386005) of \rm{NGC~5940} carried out between 2010 and 2011 for improved statistics. We grouped the data to have a minimum of 5 counts per bin for use with {\tt XSPEC} $\chi^{2}$ statistic. We started by fitting the $1.5-7.0\,\mathrm{keV}$ spectra with a simple powerlaw. When extrapolated down to $0.3\,\mathrm{keV}$, the presence of strong soft X-ray excess was confirmed to be evident in the spectra  as shown in Fig. \ref{swift}. However, both the powerlaw photon index $\Gamma$ and {\tt tbabs} column density $N_{H}$ could not be constrained due to the poorer data quality when compared to \textit{XMM-Newton}.
\section{Discussion}
We carried out spectral analysis on the $2012$ \textit{XMM-Newton} observation of the Seyfert 1 AGN \rm{NGC~5940}. The data revealed the presence of strong soft X-ray emission in the source below $2\,\mathrm{keV}$ in excess of the dominant X-ray powerlaw. Here, we investigated the possible origin of this emission component. 

We started by fitting a phenomenological model to the data, namely; the multicolour disc blackbody, to model the soft X-ray excess with the addition of the powerlaw model for the X-ray emission above $\sim2\,\mathrm{keV}$. As is the case with most fits using the blackbody model, the inferred temperature of $\sim0.18\,\mathrm{keV}$ is too high for an accretion disc around a blackhole mass of $\sim10^{7}\,M_{\odot}$. Standard accretion disc theory predicts that the inner disc temperature of such a system is given by the relation
\begin{small}
\begin{equation*}
T_{eff}\sim6.3\times10^{5}\left(\dfrac{\dot{M}}{\dot{M}_{Edd}}\right)^{1/4}\left(\dfrac{M}{10^{8}M_{\odot}}\right)^{-1/4}\left(\dfrac{R}{R_{s}}\right)^{-3/4},
\end{equation*}
\end{small}
where $\dot{M}/\dot{M}_{Edd}$ is the Eddington-scaled accretion rate, $M$ is the blackhole mass, $R$ and $R_{s}$ are respectively the disc annular radius and the Schwarzschild radius. Thus, for a Schwarzschild blackhole of $\sim10^{7}\,M_{\odot}$, at the inner edge $T_{eff}$ should be of order $~30\,\mathrm{eV}$ \citep[e.g.,][]{2014MNRAS.440..106B, 2017MNRAS.466.3951A}. We subsequently modelled the lower X-ray energy features using the ionised partial covering model -- multiplied by the powerlaw model. This model posits that the lower energy features may include contributions from partial covering of the X-ray source due to partially ionised material along the line of sight \citep[e.g.,][]{2006A&A...453L..13M, 2008MNRAS.385L.108R}. This model provided a very good fit to the data and has been employed to study the soft X-ray excess and the spectral variability in \rm{PG~0844+349} \citep{2011MNRAS.412..161G} and \rm{Zw~229.015} \citep{2019MNRAS.488.4831T} among others with similarly good fits. However, a compact absorber would imply the presence of a strong iron fluorescence line which is not observed in the data. It is plausible that the geometry of the absorber may be responsible for the non-detection of the line. Longer and possibly joint multiwavelength observations of the source will throw more light on this. 

We then applied two more physically motivated models -- the warm Comptonisation and the blurred reflection models -- that have been applied to model successfully the soft excess in several Seyfert 1 AGN e.g., \rm{1H~0707-495} \citep{2009Natur.459..540F}, \rm{Mrk~509} \citep{2011A&A...534A..39M}, \rm{ESO~113-G010} \citep{2013ApJ...764L...9C}, \rm{II~Zw~177} \citep{2016MNRAS.457..875P}, \rm{Ton~S180} \citep{2020MNRAS.497.2352M}. Both the thermal Comptonisation and the blurred reflection models provided acceptable fit to the data implying they both explained the origin of the soft excess.

The warm Comptonisation model ({\tt optxagnf}) considers the soft X-ray excess to originate due to repeated inverse-Compton scattering of seed disc photons in a warm, optically thick corona. The fitted optical depth $\tau$ and temperature $kT_{SE}$ of the warm corona were respectively $\sim13$ and $0.4\,\mathrm{keV}$ with a moderately high accretion rate value of about $5\%$ of Eddington. This inferred value of accretion rate supports earlier analysis carried out on the source \citep{2016MNRAS.458.1839C}. The model favoured a maximally spinning blackhole with corona radius, photon index and power fraction of the powerlaw having values $\sim18\,R_{g}$, $\sim1.7$ and $77\%$ respectively. The modelled $0.3-10\,\mathrm{keV}$ flux and luminosity came out to be $8.2\times10^{-12}\,\mathrm{ergs\,cm^{-2}\,s^{-1}}$ and $2.2\times10^{43}\,\mathrm{ergs\,s^{-1}}$ respectively.

The blurred reflection model ({\tt relxill}) on the other hand considers that strong X-ray reflection in the inner region of the accretion disc, blurred and smoothed due to strong general relativistic effects in this region is responsible for the soft X-ray excess emission. Fit with the {\tt relxillcp} flavour of the model gave reflection fraction value of $\sim0.5$ implying moderately strong X-ray reflection in the disc innermost region. The model equally favoured a maximally spinning blackhole having photon index $\sim1.9$. The {\tt relxill+nthcomp} version of the model provided a very good fit to the data with photon index value of $\sim2$. The value of the ionisation parameter here is much smaller at log$\epsilon=0.02$. An additional evidence for blurred inner disc reflection origin for the soft X-ray excess is the expected presence of broad iron $K_\alpha$ line at $\sim6.4\,\mathrm{keV}$ as this line is also believed to be produced due to strong X-ray reflection in the disc \citep{1995Natur.375..659T}. The line is however not detectable in this observation -- to any level of significance. Moreso, the RGS spectra with its higher resolution should normally reveal some of the prominent line features that blend to produce the soft excess, but the RGS data of this source appear to be largely featureless, fitted by a simple powerlaw. Longer focused observation in the X-rays and optical/UV bands may be required to probe this further.
\section{Conclusion \& Summary}
We studied the spectral properties of the nearby Seyfert 1 AGN \rm{NGC~5940} using predominantly its 2012 \textit{XMM-Newton} observation supplemented with data from \textit{Swift}. This source has been spared of scrutiny over the past years despite its proximity and availability of usable data. The data revealed strong soft X-ray excess emission below $\sim2\,\mathrm{keV}$ as is commonly seen in many unobscured AGN. Using state-of-the-art spectral models, we went ahead to probe the origin of the soft X-ray excess emission.

The phenomenological model applied provided acceptable fits to the \textit{XMM-Newton} EPIC-pn data of the source. From the disc blackbody plus powerlaw model, the model-predicted temperature is much higher than expected for accretion discs around supermassive blackholes. The model for partial covering from partially ionised material tend to provide a very good fit to the data for the origin of the soft excess. However, longer duration, high resolution multiwavelength observations will be required to probe further the possibility of partial covering in the source. Such observations will better probe additional emission/absorption and other signatures that should accompany an ionised partial covering scenario

Two additional physically motivated models were also applied -- The thermal Comptonisation and the blurred reflection models. We found that both models provided acceptable fits and thus can explain the origin of the soft X-ray excess emission well although with suggestive preference for the thermal Comptonisation model. This is because other features that should normally accompany strong reflection -- like the broad iron $K_{\alpha}$ line as well as emission line features in the RGS spectra -- were not observed in our analysis. 

Timing analysis may provide useful insight into understanding the importance of thermal Componisation and/or blurred reflection in the accretion flow physics of the source. For example, if cross-correlation analysis reveals that optical/UV disc photons lead the X-rays in their variability, then thermal Comptonisation could play a dominant role in the origin of the soft excess \citep{2019ApJ...870L..13A}. If on the contrary, X-rays lead, then reprocessing and therefore blurred reflection may play a more dominant role \citep{1991ApJ...371..541K, 2007MNRAS.380..669C}. It is equally possible that both thermal Comptonisation and relativistically blurred reflection play a role in the overall spectral energy distribution of the source such that while warm Comptonisation explains the origin of the soft X-ray excess, reflection may explain the line features as well as the hard X-ray excess. Simultaneous observations of the source at higher X-ray energies with instruments like the \textit{NuSTAR} space telescope and at lower optical/UV energies will be crucial for a more detailed probe of the spectral behaviour of this peculiar source.

\acknowledgments
This work made use of data supplied by the XMM-Newton Science Archive and the UK Swift Science Data Centre at the University of Leicester. The anonymous referee is appreciated for useful comments that improved the quality of the manuscript.

\section*{Additional Statements}



\begin{fundinginformation}
None
\end{fundinginformation}

\begin{dataavailability}
Details of all the data analysed in this study are included in the article.
\end{dataavailability}



\begin{ethics}
\begin{conflict}
The author has no competing interests to declare regarding the content of this article.\\\\
{\bf Informed consent} None
\end{conflict}
\end{ethics}

%
%
%
%
%
%
%
%
%
%
%

\bibliographystyle{spr-mp-nameyear-cnd}
\bibliography{bibtex}

\begin{thebibliography}{58}
\ifx \bisbn   \undefined \def \bisbn  #1{ISBN #1}\fi
\ifx \binits  \undefined \def \binits#1{#1}\fi
\ifx \bauthor  \undefined \def \bauthor#1{#1}\fi
\ifx \batitle  \undefined \def \batitle#1{#1}\fi
\ifx \bjtitle  \undefined \def \bjtitle#1{#1}\fi
\ifx \bvolume  \undefined \def \bvolume#1{\textbf{#1}}\fi
\ifx \byear  \undefined \def \byear#1{#1}\fi
\ifx \bissue  \undefined \def \bissue#1{#1}\fi
\ifx \bfpage  \undefined \def \bfpage#1{#1}\fi
\ifx \blpage  \undefined \def \blpage #1{#1}\fi
\ifx \burl  \undefined \def \burl#1{\textsf{#1}}\fi
\ifx \doiurl  \undefined \def \doiurl#1{\textsf{#1}}\fi
\ifx \betal  \undefined \def \betal{\textit{et al.}}\fi
\ifx \binstitute  \undefined \def \binstitute#1{#1}\fi
\ifx \binstitutionaled  \undefined \def \binstitutionaled#1{#1}\fi
\ifx \bctitle  \undefined \def \bctitle#1{#1}\fi
\ifx \beditor  \undefined \def \beditor#1{#1}\fi
\ifx \bpublisher  \undefined \def \bpublisher#1{#1}\fi
\ifx \bbtitle  \undefined \def \bbtitle#1{#1}\fi
\ifx \bedition  \undefined \def \bedition#1{#1}\fi
\ifx \bseriesno  \undefined \def \bseriesno#1{#1}\fi
\ifx \blocation  \undefined \def \blocation#1{#1}\fi
\ifx \bsertitle  \undefined \def \bsertitle#1{#1}\fi
\ifx \bsnm \undefined \def \bsnm#1{#1}\fi
\ifx \bsuffix \undefined \def \bsuffix#1{#1}\fi
\ifx \bparticle \undefined \def \bparticle#1{#1}\fi
\ifx \barticle \undefined \def \barticle#1{#1}\fi
\ifx \bconfdate \undefined \def \bconfdate #1{#1} \fi
\ifx \botherref \undefined \def \botherref #1{#1} \fi
\ifx \url \undefined \def \url#1{\textsf{#1}} \fi
\ifx \bchapter \undefined \def \bchapter#1{#1} \fi
\ifx \bbook \undefined \def \bbook#1{#1} \fi
\ifx \bcomment \undefined \def \bcomment#1{#1} \fi
\ifx \oauthor \undefined \def \oauthor#1{#1} \fi
\ifx \citeauthoryear \undefined \def \citeauthoryear#1{#1} \fi
\ifx \endbibitem  \undefined \def \endbibitem {}\fi
\ifx \bconflocation  \undefined \def \bconflocation#1{#1} \fi
\ifx \arxivurl  \undefined \def \arxivurl#1{\textsf{#1}} \fi
\csname PreBibitemsHook\endcsname

\bibitem[\protect\citeauthoryear{{Adegoke} et~al.}{2017}]{2017MNRAS.466.3951A}
\begin{barticle}
\bauthor{\bsnm{{Adegoke}}, \binits{O.}},
\bauthor{\bsnm{{Rakshit}}, \binits{S.}},
\bauthor{\bsnm{{Mukhopadhyay}}, \binits{B.}}:
\bjtitle{\mnras}
\bvolume{466},
\bfpage{3951}
(\byear{2017}).
\doiurl{https://doi.org/10.1093
/mnras/stw3320}.
\arxivurl{1612.06817}
\end{barticle}
\endbibitem

\bibitem[\protect\citeauthoryear{{Adegoke} et~al.}{2019}]{2019ApJ...870L..13A}
\begin{barticle}
\bauthor{\bsnm{{Adegoke}}, \binits{O.}},
\bauthor{\bsnm{{Dewangan}}, \binits{G.C.}},
\bauthor{\bsnm{{Pawar}}, \binits{P.}},
\bauthor{\bsnm{{Pal}}, \binits{M.}}:
\bjtitle{\apjl}
\bvolume{870}(\bissue{2}),
\bfpage{13}
(\byear{2019}).
\doiurl{https://doi.org/10.3847
/2041-8213/aaf8ab}.
\arxivurl{1812.08056}
\end{barticle}
\endbibitem

\bibitem[\protect\citeauthoryear{{Arnaud}}{1996}]{1996ASPC..101...17A}
\begin{bchapter}
\bauthor{\bsnm{{Arnaud}}, \binits{K.A.}}:
In: \beditor{\bsnm{{Jacoby}}, \binits{G.H.}},
\beditor{\bsnm{{Barnes}}, \binits{J.}} (eds.)
\bbtitle{Astronomical Data Analysis Software and Systems V}.
\bsertitle{Astronomical Society of the Pacific Conference Series},
vol. \bseriesno{101},
p. \bfpage{17}
(\byear{1996})
\end{bchapter}
\endbibitem

\bibitem[\protect\citeauthoryear{{Arnaud} et~al.}{1985}]{1985MNRAS.217..105A}
\begin{barticle}
\bauthor{\bsnm{{Arnaud}}, \binits{K.A.}},
\bauthor{\bsnm{{Branduardi-Raymont}}, \binits{G.}},
\bauthor{\bsnm{{Culhane}}, \binits{J.L.}},
\bauthor{\bsnm{{Fabian}}, \binits{A.C.}},
\bauthor{\bsnm{{Hazard}}, \binits{C.}},
\bauthor{\bsnm{{McGlynn}}, \binits{T.A.}},
\bauthor{\bsnm{{Shafer}}, \binits{R.A.}},
\bauthor{\bsnm{{Tennant}}, \binits{A.F.}},
\bauthor{\bsnm{{Ward}}, \binits{M.J.}}:
\bjtitle{\mnras}
\bvolume{217},
\bfpage{105}
(\byear{1985}).
\doiurl{https://doi.org/10.1093/mnras/217.1.105}
\end{barticle}
\endbibitem

\bibitem[\protect\citeauthoryear{{Barth} et~al.}{2013}]{2013ApJ...769..128B}
\begin{barticle}
\bauthor{\bsnm{{Barth}}, \binits{A.J.}},
\bauthor{\bsnm{{Pancoast}}, \binits{A.}},
\bauthor{\bsnm{{Bennert}}, \binits{V.N.}},
\bauthor{\bsnm{{Brewer}}, \binits{B.J.}},
\bauthor{\bsnm{{Canalizo}}, \binits{G.}},
\bauthor{\bsnm{{Filippenko}}, \binits{A.V.}},
\bauthor{\bsnm{{Gates}}, \binits{E.L.}},
\bauthor{\bsnm{{Greene}}, \binits{J.E.}},
\bauthor{\bsnm{{Li}}, \binits{W.}},
\bauthor{\bsnm{{Malkan}}, \binits{M.A.}},
\bauthor{\bsnm{{Sand}}, \binits{D.J.}},
\bauthor{\bsnm{{Stern}}, \binits{D.}},
\bauthor{\bsnm{{Treu}}, \binits{T.}},
\bauthor{\bsnm{{Woo}}, \binits{J.-H.}},
\bauthor{\bsnm{{Assef}}, \binits{R.J.}},
\bauthor{\bsnm{{Bae}}, \binits{H.-J.}},
\bauthor{\bsnm{{Buehler}}, \binits{T.}},
\bauthor{\bsnm{{Cenko}}, \binits{S.B.}},
\bauthor{\bsnm{{Clubb}}, \binits{K.I.}},
\bauthor{\bsnm{{Cooper}}, \binits{M.C.}},
\bauthor{\bsnm{{Diamond-Stanic}}, \binits{A.M.}},
\bauthor{\bsnm{{H{\"o}nig}}, \binits{S.F.}},
\bauthor{\bsnm{{Joner}}, \binits{M.D.}},
\bauthor{\bsnm{{Laney}}, \binits{C.D.}},
\bauthor{\bsnm{{Lazarova}}, \binits{M.S.}},
\bauthor{\bsnm{{Nierenberg}}, \binits{A.M.}},
\bauthor{\bsnm{{Silverman}}, \binits{J.M.}},
\bauthor{\bsnm{{Tollerud}}, \binits{E.J.}},
\bauthor{\bsnm{{Walsh}}, \binits{J.L.}}:
\bjtitle{\apj}
\bvolume{769}(\bissue{2}),
\bfpage{128}
(\byear{2013}).
\doiurl{https://doi.org/10.1088/0004-637X/769/2/128}.
\arxivurl{1304.4643}
\end{barticle}
\endbibitem

\bibitem[\protect\citeauthoryear{{Barth} et~al.}{2015}]{2015ApJS..217...26B}
\begin{barticle}
\bauthor{\bsnm{{Barth}}, \binits{A.J.}},
\bauthor{\bsnm{{Bennert}}, \binits{V.N.}},
\bauthor{\bsnm{{Canalizo}}, \binits{G.}},
\bauthor{\bsnm{{Filippenko}}, \binits{A.V.}},
\bauthor{\bsnm{{Gates}}, \binits{E.L.}},
\bauthor{\bsnm{{Greene}}, \binits{J.E.}},
\bauthor{\bsnm{{Li}}, \binits{W.}},
\bauthor{\bsnm{{Malkan}}, \binits{M.A.}},
\bauthor{\bsnm{{Pancoast}}, \binits{A.}},
\bauthor{\bsnm{{Sand}}, \binits{D.J.}},
\bauthor{\bsnm{{Stern}}, \binits{D.}},
\bauthor{\bsnm{{Treu}}, \binits{T.}},
\bauthor{\bsnm{{Woo}}, \binits{J.-H.}},
\bauthor{\bsnm{{Assef}}, \binits{R.J.}},
\bauthor{\bsnm{{Bae}}, \binits{H.-J.}},
\bauthor{\bsnm{{Brewer}}, \binits{B.J.}},
\bauthor{\bsnm{{Cenko}}, \binits{S.B.}},
\bauthor{\bsnm{{Clubb}}, \binits{K.I.}},
\bauthor{\bsnm{{Cooper}}, \binits{M.C.}},
\bauthor{\bsnm{{Diamond-Stanic}}, \binits{A.M.}},
\bauthor{\bsnm{{Hiner}}, \binits{K.D.}},
\bauthor{\bsnm{{H{\"o}nig}}, \binits{S.F.}},
\bauthor{\bsnm{{Hsiao}}, \binits{E.}},
\bauthor{\bsnm{{Kandrashoff}}, \binits{M.T.}},
\bauthor{\bsnm{{Lazarova}}, \binits{M.S.}},
\bauthor{\bsnm{{Nierenberg}}, \binits{A.M.}},
\bauthor{\bsnm{{Rex}}, \binits{J.}},
\bauthor{\bsnm{{Silverman}}, \binits{J.M.}},
\bauthor{\bsnm{{Tollerud}}, \binits{E.J.}},
\bauthor{\bsnm{{Walsh}}, \binits{J.L.}}:
\bjtitle{\apjs}
\bvolume{217}(\bissue{2}),
\bfpage{26}
(\byear{2015}).
\doiurl{https://doi.org/10.1088/0067-0049/217/2/26}.
\arxivurl{1503.01146}
\end{barticle}
\endbibitem

\bibitem[\protect\citeauthoryear{{Bentz} and
  {Katz}}{2015}]{2015PASP..127...67B}
\begin{barticle}
\bauthor{\bsnm{{Bentz}}, \binits{M.C.}},
\bauthor{\bsnm{{Katz}}, \binits{S.}}:
\bjtitle{\pasp}
\bvolume{127}(\bissue{947}),
\bfpage{67}
(\byear{2015}).
\doiurl{https://doi.org/10.1086/679601}.
\arxivurl{1411.2596}
\end{barticle}
\endbibitem

\bibitem[\protect\citeauthoryear{{Bhattacharyya}
  et~al.}{2014}]{2014MNRAS.440..106B}
\begin{barticle}
\bauthor{\bsnm{{Bhattacharyya}}, \binits{S.}},
\bauthor{\bsnm{{Bhatt}}, \binits{H.}},
\bauthor{\bsnm{{Bhatt}}, \binits{N.}},
\bauthor{\bsnm{{Singh}}, \binits{K.K.}}:
\bjtitle{\mnras}
\bvolume{440}(\bissue{1}),
\bfpage{106}
(\byear{2014}).
\doiurl{https://doi.org
/10.1093/mnras/stu239}.
\arxivurl{1301.1150}
\end{barticle}
\endbibitem

\bibitem[\protect\citeauthoryear{{Burrows} et~al.}{2005}]{2005SSRv..120..165B}
\begin{barticle}
\bauthor{\bsnm{{Burrows}}, \binits{D.N.}},
\bauthor{\bsnm{{Hill}}, \binits{J.E.}},
\bauthor{\bsnm{{Nousek}}, \binits{J.A.}},
\bauthor{\bsnm{{Kennea}}, \binits{J.A.}},
\bauthor{\bsnm{{Wells}}, \binits{A.}},
\bauthor{\bsnm{{Osborne}}, \binits{J.P.}},
\bauthor{\bsnm{{Abbey}}, \binits{A.F.}},
\bauthor{\bsnm{{Beardmore}}, \binits{A.}},
\bauthor{\bsnm{{Mukerjee}}, \binits{K.}},
\bauthor{\bsnm{{Short}}, \binits{A.D.T.}},
\bauthor{\bsnm{{Chincarini}}, \binits{G.}},
\bauthor{\bsnm{{Campana}}, \binits{S.}},
\bauthor{\bsnm{{Citterio}}, \binits{O.}},
\bauthor{\bsnm{{Moretti}}, \binits{A.}},
\bauthor{\bsnm{{Pagani}}, \binits{C.}},
\bauthor{\bsnm{{Tagliaferri}}, \binits{G.}},
\bauthor{\bsnm{{Giommi}}, \binits{P.}},
\bauthor{\bsnm{{Capalbi}}, \binits{M.}},
\bauthor{\bsnm{{Tamburelli}}, \binits{F.}},
\bauthor{\bsnm{{Angelini}}, \binits{L.}},
\bauthor{\bsnm{{Cusumano}}, \binits{G.}},
\bauthor{\bsnm{{Br{\"a}uninger}}, \binits{H.W.}},
\bauthor{\bsnm{{Burkert}}, \binits{W.}},
\bauthor{\bsnm{{Hartner}}, \binits{G.D.}}:
\bjtitle{\ssr}
\bvolume{120}(\bissue{3-4}),
\bfpage{165}
(\byear{2005}).
\doiurl{https://doi.org/10.1007/s11214-005-5097-2}.
\arxivurl{astro-ph/0508071}
\end{barticle}
\endbibitem

\bibitem[\protect\citeauthoryear{{Cackett} et~al.}{2007}]{2007MNRAS.380..669C}
\begin{barticle}
\bauthor{\bsnm{{Cackett}}, \binits{E.M.}},
\bauthor{\bsnm{{Horne}}, \binits{K.}},
\bauthor{\bsnm{{Winkler}}, \binits{H.}}:
\bjtitle{\mnras}
\bvolume{380},
\bfpage{669}
(\byear{2007}).
\doiurl{https://doi.org/10.1111
/j.1365-2966.2007.12098.x}.
\arxivurl{0706.1464}
\end{barticle}
\endbibitem

\bibitem[\protect\citeauthoryear{{Cackett} et~al.}{2013}]{2013ApJ...764L...9C}
\begin{barticle}
\bauthor{\bsnm{{Cackett}}, \binits{E.M.}},
\bauthor{\bsnm{{Fabian}}, \binits{A.C.}},
\bauthor{\bsnm{{Zogbhi}}, \binits{A.}},
\bauthor{\bsnm{{Kara}}, \binits{E.}},
\bauthor{\bsnm{{Rey-
nolds}}, \binits{C.}},
\bauthor{\bsnm{{Uttley}}, \binits{P.}}:
\bjtitle{\apjl}
\bvolume{764}(\bissue{1}),
\bfpage{9}
(\byear{2013}).
\doiurl{https://doi.org/10.1088/2041-8205/764/1/L9}.
\arxivurl{1210.7874}
\end{barticle}
\endbibitem

\bibitem[\protect\citeauthoryear{{Castell{\'o}-Mor}
  et~al.}{2016}]{2016MNRAS.458.1839C}
\begin{barticle}
\bauthor{\bsnm{{Castell{\'o}-Mor}}, \binits{N.}},
\bauthor{\bsnm{{Netzer}}, \binits{H.}},
\bauthor{\bsnm{{Kaspi}}, \binits{S.}}:
\bjtitle{\mnras}
\bvolume{458}(\bissue{2}),
\bfpage{1839}
(\byear{2016}).
\doiurl{https://doi.org/10.1093
/mnras/stw445}.
\arxivurl{1601.07177}
\end{barticle}
\endbibitem

\bibitem[\protect\citeauthoryear{{Crummy} et~al.}{2005}]{2005MNRAS.361.1197C}
\begin{barticle}
\bauthor{\bsnm{{Crummy}}, \binits{J.}},
\bauthor{\bsnm{{Fabian}}, \binits{A.C.}},
\bauthor{\bsnm{{Brandt}}, \binits{W.N.}},
\bauthor{\bsnm{{Boller}}, \binits{T.}}:
\bjtitle{\mnras}
\bvolume{361}(\bissue{4}),
\bfpage{1197}
(\byear{2005}).
\doiurl{https://doi.org
/10.1111/j.1365-2966.2005.09258.x}.
\arxivurl{astro-ph/0506119}
\end{barticle}
\endbibitem

\bibitem[\protect\citeauthoryear{{Crummy} et~al.}{2006}]{2006MNRAS.365.1067C}
\begin{barticle}
\bauthor{\bsnm{{Crummy}}, \binits{J.}},
\bauthor{\bsnm{{Fabian}}, \binits{A.C.}},
\bauthor{\bsnm{{Gallo}}, \binits{L.}},
\bauthor{\bsnm{{Ross}}, \binits{R.R.}}:
\bjtitle{\mnras}
\bvolume{365}(\bissue{4}),
\bfpage{1067}
(\byear{2006}).
\doiurl{https://doi.org
/10.1111/j.1365-2966.2005.09844.x}.
\arxivurl{astro-ph/0511457}
\end{barticle}
\endbibitem

\bibitem[\protect\citeauthoryear{{Dauser} et~al.}{2014}]{2014MNRAS.444L.100D}
\begin{barticle}
\bauthor{\bsnm{{Dauser}}, \binits{T.}},
\bauthor{\bsnm{{Garcia}}, \binits{J.}},
\bauthor{\bsnm{{Parker}}, \binits{M.L.}},
\bauthor{\bsnm{{Fabian}}, \binits{A.C.}},
\bauthor{\bsnm{{Wilms}}, \binits{J.}}:
\bjtitle{\mnras}
\bvolume{444},
\bfpage{100}
(\byear{2014}).
\doiurl{https://doi
.org/10.1093/mnrasl/slu125}.
\arxivurl{1408.2347}
\end{barticle}
\endbibitem

\bibitem[\protect\citeauthoryear{{den Herder}
  et~al.}{2001}]{2001A&A...365L...7D}
\begin{barticle}
\bauthor{\bsnm{{den Herder}}, \binits{J.W.}},
\bauthor{\bsnm{{Brinkman}}, \binits{A.C.}},
\bauthor{\bsnm{{Kahn}}, \binits{S.M.}},
\bauthor{\bsnm{{Bran-
duardi-Raymont}}, \binits{G.}},
\bauthor{\bsnm{{Thomsen}}, \binits{K.}},
\bauthor{\bsnm{{Aarts}}, \binits{H.}},
\bauthor{\bsnm{{Audard}}, \binits{M.}},
\bauthor{\bsnm{{Bixler}}, \binits{J.V.}},
\bauthor{\bsnm{{den Boggende}}, \binits{A.J.}},
\bauthor{\bsnm{{Cottam}}, \binits{J.}},
\bauthor{\bsnm{{Decker}}, \binits{T.}},
\bauthor{\bsnm{{Dubbeldam}}, \binits{L.}},
\bauthor{\bsnm{{Erd}}, \binits{C.}},
\bauthor{\bsnm{{Goulooze}}, \binits{H.}},
\bauthor{\bsnm{{G{\"u}del}}, \binits{M.}},
\bauthor{\bsnm{{Guttridge}}, \binits{P.}},
\bauthor{\bsnm{{Hailey}}, \binits{C.J.}},
\bauthor{\bsnm{{Janabi}}, \binits{K.A.}},
\bauthor{\bsnm{{Kaastra}}, \binits{J.S.}},
\bauthor{\bsnm{{de Korte}}, \binits{P.A.J.}},
\bauthor{\bsnm{{van Leeuwen}}, \binits{B.J.}},
\bauthor{\bsnm{{Mauche}}, \binits{C.}},
\bauthor{\bsnm{{McCalden}}, \binits{A.J.}},
\bauthor{\bsnm{{Mewe}}, \binits{R.}},
\bauthor{\bsnm{{Naber}}, \binits{A.}},
\bauthor{\bsnm{{Paerels}}, \binits{F.B.}},
\bauthor{\bsnm{{Peterson}}, \binits{J.R.}},
\bauthor{\bsnm{{Rasmussen}}, \binits{A.P.}},
\bauthor{\bsnm{{Rees}}, \binits{K.}},
\bauthor{\bsnm{{Sakelliou}}, \binits{I.}},
\bauthor{\bsnm{{Sako}}, \binits{M.}},
\bauthor{\bsnm{{Spodek}}, \binits{J.}},
\bauthor{\bsnm{{Stern}}, \binits{M.}},
\bauthor{\bsnm{{Tamura}}, \binits{T.}},
\bauthor{\bsnm{{Tandy}}, \binits{J.}},
\bauthor{\bsnm{{de Vries}}, \binits{C.P.}},
\bauthor{\bsnm{{Welch}}, \binits{S.}},
\bauthor{\bsnm{{Zehnder}}, \binits{A.}}:
\bjtitle{\aap}
\bvolume{365},
\bfpage{7}
(\byear{2001}).
\doiurl{https://doi.org/10.1051/0004-6361:20000058}
\end{barticle}
\endbibitem

\bibitem[\protect\citeauthoryear{{Done} et~al.}{2012}]{2012MNRAS.420.1848D}
\begin{barticle}
\bauthor{\bsnm{{Done}}, \binits{C.}},
\bauthor{\bsnm{{Davis}}, \binits{S.W.}},
\bauthor{\bsnm{{Jin}}, \binits{C.}},
\bauthor{\bsnm{{Blaes}}, \binits{O.}},
\bauthor{\bsnm{{Ward}}, \binits{M.}}:
\bjtitle{\mnras}
\bvolume{420}(\bissue{3}),
\bfpage{1848}
(\byear{2012}).
\doiurl{https://doi.org/
10.1111/j.1365-2966.2011.19779.x}.
\arxivurl{1107.5429}
\end{barticle}
\endbibitem

\bibitem[\protect\citeauthoryear{{Evans} et~al.}{2009}]{2009MNRAS.397.1177E}
\begin{barticle}
\bauthor{\bsnm{{Evans}}, \binits{P.A.}},
\bauthor{\bsnm{{Beardmore}}, \binits{A.P.}},
\bauthor{\bsnm{{Page}}, \binits{K.L.}},
\bauthor{\bsnm{{Osborne}}, \binits{J.P.}},
\bauthor{\bsnm{{O'Brien}}, \binits{P.T.}},
\bauthor{\bsnm{{Willingale}}, \binits{R.}},
\bauthor{\bsnm{{Starling}}, \binits{R.L.C.}},
\bauthor{\bsnm{{Burrows}}, \binits{D.N.}},
\bauthor{\bsnm{{Godet}}, \binits{O.}},
\bauthor{\bsnm{{Vetere}}, \binits{L.}},
\bauthor{\bsnm{{Racusin}}, \binits{J.}},
\bauthor{\bsnm{{Goad}}, \binits{M.R.}},
\bauthor{\bsnm{{Wiersema}}, \binits{K.}},
\bauthor{\bsnm{{Angelini}}, \binits{L.}},
\bauthor{\bsnm{{Capalbi}}, \binits{M.}},
\bauthor{\bsnm{{Chincarini}}, \binits{G.}},
\bauthor{\bsnm{{Gehrels}}, \binits{N.}},
\bauthor{\bsnm{{Kennea}}, \binits{J.A.}},
\bauthor{\bsnm{{Margutti}}, \binits{R.}},
\bauthor{\bsnm{{Morris}}, \binits{D.C.}},
\bauthor{\bsnm{{Mountford}}, \binits{C.J.}},
\bauthor{\bsnm{{Pagani}}, \binits{C.}},
\bauthor{\bsnm{{Perri}}, \binits{M.}},
\bauthor{\bsnm{{Romano}}, \binits{P.}},
\bauthor{\bsnm{{Tanvir}}, \binits{N.}}:
\bjtitle{\mnras}
\bvolume{397}(\bissue{3}),
\bfpage{1177}
(\byear{2009}).
\doiurl{https://doi.org/10.1111/j.1365-2966.2009.14913.x}.
\arxivurl{0812.3662}
\end{barticle}
\endbibitem

\bibitem[\protect\citeauthoryear{{Fabian} et~al.}{2004}]{2004MNRAS.353.1071F}
\begin{barticle}
\bauthor{\bsnm{{Fabian}}, \binits{A.C.}},
\bauthor{\bsnm{{Miniutti}}, \binits{G.}},
\bauthor{\bsnm{{Gallo}}, \binits{L.}},
\bauthor{\bsnm{{Boller}}, \binits{T.}},
\bauthor{\bsnm{{Tanaka}}, \binits{Y.}},
\bauthor{\bsnm{{Vaughan}}, \binits{S.}},
\bauthor{\bsnm{{Ross}}, \binits{R.R.}}:
\bjtitle{\mnras}
\bvolume{353}(\bissue{4}),
\bfpage{1071}
(\byear{2004}).
\doiurl{https://doi.org/10.1111/j.1365-2966.2004.08036.x}.
\arxivurl{astro-ph/0405160}
\end{barticle}
\endbibitem

\bibitem[\protect\citeauthoryear{{Fabian} et~al.}{2009}]{2009Natur.459..540F}
\begin{barticle}
\bauthor{\bsnm{{Fabian}}, \binits{A.C.}},
\bauthor{\bsnm{{Zoghbi}}, \binits{A.}},
\bauthor{\bsnm{{Ross}}, \binits{R.R.}},
\bauthor{\bsnm{{Uttley}}, \binits{P.}},
\bauthor{\bsnm{{Gallo}}, \binits{L.C.}},
\bauthor{\bsnm{{Brandt}}, \binits{W.N.}},
\bauthor{\bsnm{{Blustin}}, \binits{A.J.}},
\bauthor{\bsnm{{Boller}}, \binits{T.}},
\bauthor{\bsnm{{Caballero-Garcia}}, \binits{M.D.}},
\bauthor{\bsnm{{Larsson}}, \binits{J.}},
\bauthor{\bsnm{{Miller}}, \binits{J.M.}},
\bauthor{\bsnm{{Miniutti}}, \binits{G.}},
\bauthor{\bsnm{{Ponti}}, \binits{G.}},
\bauthor{\bsnm{{Reis}}, \binits{R.C.}},
\bauthor{\bsnm{{Reynolds}}, \binits{C.S.}},
\bauthor{\bsnm{{Tanaka}}, \binits{Y.}},
\bauthor{\bsnm{{Young}}, \binits{A.J.}}:
\bjtitle{\nat}
\bvolume{459}(\bissue{7246}),
\bfpage{540}
(\byear{2009}).
\doiurl{https://doi.org/10.1038/na-
ture08007}
\end{barticle}
\endbibitem

\bibitem[\protect\citeauthoryear{{Gallo} et~al.}{2011}]{2011MNRAS.412..161G}
\begin{barticle}
\bauthor{\bsnm{{Gallo}}, \binits{L.C.}},
\bauthor{\bsnm{{Grupe}}, \binits{D.}},
\bauthor{\bsnm{{Schartel}}, \binits{N.}},
\bauthor{\bsnm{{Komossa}}, \binits{S.}},
\bauthor{\bsnm{{Miniutti}}, \binits{G.}},
\bauthor{\bsnm{{Fabian}}, \binits{A.C.}},
\bauthor{\bsnm{{Santos-Lleo}}, \binits{M.}}:
\bjtitle{\mnras}
\bvolume{412}(\bissue{1}),
\bfpage{161}
(\byear{2011}).
\doiurl{https://doi.org/10.1111/j.1365-2966.2010.17894.x}.
\arxivurl{1010.4453}
\end{barticle}
\endbibitem

\bibitem[\protect\citeauthoryear{{Garc{\'\i}a}
  et~al.}{2014}]{2014ApJ...782...76G}
\begin{barticle}
\bauthor{\bsnm{{Garc{\'\i}a}}, \binits{J.}},
\bauthor{\bsnm{{Dauser}}, \binits{T.}},
\bauthor{\bsnm{{Lohfink}}, \binits{A.}},
\bauthor{\bsnm{{Kallman}}, \binits{T.R.}},
\bauthor{\bsnm{{Steiner}}, \binits{J.F.}},
\bauthor{\bsnm{{McClintock}}, \binits{J.E.}},
\bauthor{\bsnm{{Brenneman}}, \binits{L.}},
\bauthor{\bsnm{{Wilms}}, \binits{J.}},
\bauthor{\bsnm{{Eikmann}}, \binits{W.}},
\bauthor{\bsnm{{Reynolds}}, \binits{C.S.}},
\bauthor{\bsnm{{Tombesi}}, \binits{F.}}:
\bjtitle{\apj}
\bvolume{782}(\bissue{2}),
\bfpage{76}
(\byear{2014}).
\doiurl{https://doi.org/10.1088/0004-637X/782/2/76}.
\arxivurl{1312.3231}
\end{barticle}
\endbibitem

\bibitem[\protect\citeauthoryear{{Gehrels} et~al.}{2004}]{2004ApJ...611.1005G}
\begin{barticle}
\bauthor{\bsnm{{Gehrels}}, \binits{N.}},
\bauthor{\bsnm{{Chincarini}}, \binits{G.}},
\bauthor{\bsnm{{Giommi}}, \binits{P.}},
\bauthor{\bsnm{{Mason}}, \binits{K.O.}},
\bauthor{\bsnm{{Nousek}}, \binits{J.A.}},
\bauthor{\bsnm{{Wells}}, \binits{A.A.}},
\bauthor{\bsnm{{White}}, \binits{N.E.}},
\bauthor{\bsnm{{Barthelmy}}, \binits{S.D.}},
\bauthor{\bsnm{{Burrows}}, \binits{D.N.}},
\bauthor{\bsnm{{Cominsky}}, \binits{L.R.}},
\bauthor{\bsnm{{Hurley}}, \binits{K.C.}},
\bauthor{\bsnm{{Marshall}}, \binits{F.E.}},
\bauthor{\bsnm{{M{\'e}sz{\'a}ros}}, \binits{P.}},
\bauthor{\bsnm{{Roming}}, \binits{P.W.A.}},
\bauthor{\bsnm{{Angelini}}, \binits{L.}},
\bauthor{\bsnm{{Barbier}}, \binits{L.M.}},
\bauthor{\bsnm{{Belloni}}, \binits{T.}},
\bauthor{\bsnm{{Campana}}, \binits{S.}},
\bauthor{\bsnm{{Caraveo}}, \binits{P.A.}},
\bauthor{\bsnm{{Chester}}, \binits{M.M.}},
\bauthor{\bsnm{{Citterio}}, \binits{O.}},
\bauthor{\bsnm{{Cline}}, \binits{T.L.}},
\bauthor{\bsnm{{Cropper}}, \binits{M.S.}},
\bauthor{\bsnm{{Cummings}}, \binits{J.R.}},
\bauthor{\bsnm{{Dean}}, \binits{A.J.}},
\bauthor{\bsnm{{Feigelson}}, \binits{E.D.}},
\bauthor{\bsnm{{Fenimore}}, \binits{E.E.}},
\bauthor{\bsnm{{Frail}}, \binits{D.A.}},
\bauthor{\bsnm{{Fruchter}}, \binits{A.S.}},
\bauthor{\bsnm{{Garmire}}, \binits{G.P.}},
\bauthor{\bsnm{{Gendreau}}, \binits{K.}},
\bauthor{\bsnm{{Ghisellini}}, \binits{G.}},
\bauthor{\bsnm{{Greiner}}, \binits{J.}},
\bauthor{\bsnm{{Hill}}, \binits{J.E.}},
\bauthor{\bsnm{{Hunsberger}}, \binits{S.D.}},
\bauthor{\bsnm{{Krimm}}, \binits{H.A.}},
\bauthor{\bsnm{{Kulkarni}}, \binits{S.R.}},
\bauthor{\bsnm{{Kumar}}, \binits{P.}},
\bauthor{\bsnm{{Lebrun}}, \binits{F.}},
\bauthor{\bsnm{{Lloyd-Ronning}}, \binits{N.M.}},
\bauthor{\bsnm{{Markwardt}}, \binits{C.B.}},
\bauthor{\bsnm{{Mattson}}, \binits{B.J.}},
\bauthor{\bsnm{{Mushotzky}}, \binits{R.F.}},
\bauthor{\bsnm{{Norris}}, \binits{J.P.}},
\bauthor{\bsnm{{Osborne}}, \binits{J.}},
\bauthor{\bsnm{{Paczynski}}, \binits{B.}},
\bauthor{\bsnm{{Palmer}}, \binits{D.M.}},
\bauthor{\bsnm{{Park}}, \binits{H.-S.}},
\bauthor{\bsnm{{Parsons}}, \binits{A.M.}},
\bauthor{\bsnm{{Paul}}, \binits{J.}},
\bauthor{\bsnm{{Rees}}, \binits{M.J.}},
\bauthor{\bsnm{{Reynolds}}, \binits{C.S.}},
\bauthor{\bsnm{{Rhoads}}, \binits{J.E.}},
\bauthor{\bsnm{{Sasseen}}, \binits{T.P.}},
\bauthor{\bsnm{{Schaefer}}, \binits{B.E.}},
\bauthor{\bsnm{{Short}}, \binits{A.T.}},
\bauthor{\bsnm{{Smale}}, \binits{A.P.}},
\bauthor{\bsnm{{Smith}}, \binits{I.A.}},
\bauthor{\bsnm{{Stella}}, \binits{L.}},
\bauthor{\bsnm{{Tagliaferri}}, \binits{G.}},
\bauthor{\bsnm{{Takahashi}}, \binits{T.}},
\bauthor{\bsnm{{Tashiro}}, \binits{M.}},
\bauthor{\bsnm{{Townsley}}, \binits{L.K.}},
\bauthor{\bsnm{{Tueller}}, \binits{J.}},
\bauthor{\bsnm{{Turner}}, \binits{M.J.L.}},
\bauthor{\bsnm{{Vietri}}, \binits{M.}},
\bauthor{\bsnm{{Voges}}, \binits{W.}},
\bauthor{\bsnm{{Ward}}, \binits{M.J.}},
\bauthor{\bsnm{{Willingale}}, \binits{R.}},
\bauthor{\bsnm{{Zerbi}}, \binits{F.M.}},
\bauthor{\bsnm{{Zhang}}, \binits{W.W.}}:
\bjtitle{\apj}
\bvolume{611}(\bissue{2}),
\bfpage{1005}
(\byear{2004}).
\doiurl{https://doi.org/10.1086/422091}.
\arxivurl{astro-ph/0405233}
\end{barticle}
\endbibitem

\bibitem[\protect\citeauthoryear{{Gierli{\'n}ski} and
  {Done}}{2004}]{2004MNRAS.349L...7G}
\begin{barticle}
\bauthor{\bsnm{{Gierli{\'n}ski}}, \binits{M.}},
\bauthor{\bsnm{{Done}}, \binits{C.}}:
\bjtitle{\mnras}
\bvolume{349}(\bissue{1}),
\bfpage{7}
(\byear{2004}).
\doiurl{https://doi.org/10.1111/j.1365-2966.2004.07687
.x}.
\arxivurl{astro-ph/0312271}
\end{barticle}
\endbibitem

\bibitem[\protect\citeauthoryear{{Gliozzi} and
  {Williams}}{2020}]{2020MNRAS.491..532G}
\begin{barticle}
\bauthor{\bsnm{{Gliozzi}}, \binits{M.}},
\bauthor{\bsnm{{Williams}}, \binits{J.K.}}:
\bjtitle{\mnras}
\bvolume{491}(\bissue{1}),
\bfpage{532}
(\byear{2020}).
\doiurl{https://doi.org/10.1093/mnras/stz30
05}.
\arxivurl{1910.12115}
\end{barticle}
\endbibitem

\bibitem[\protect\citeauthoryear{{Haardt} and
  {Maraschi}}{1993}]{1993ApJ...413..507H}
\begin{barticle}
\bauthor{\bsnm{{Haardt}}, \binits{F.}},
\bauthor{\bsnm{{Maraschi}}, \binits{L.}}:
\bjtitle{\apj}
\bvolume{413},
\bfpage{507}
(\byear{1993}).
\doiurl{https://doi.org/10.1086/173020}
\end{barticle}
\endbibitem

\bibitem[\protect\citeauthoryear{{HI4PI Collaboration}
  et~al.}{2016}]{2016A&A...594A.116H}
\begin{barticle}
\bauthor{\bsnm{{HI4PI Collaboration}}},
\bauthor{\bsnm{{Ben Bekhti}}, \binits{N.}},
\bauthor{\bsnm{{Fl{\"o}er}}, \binits{L.}},
\bauthor{\bsnm{{Keller}}, \binits{R.}},
\bauthor{\bsnm{{Kerp}}, \binits{J.}},
\bauthor{\bsnm{{Lenz}}, \binits{D.}},
\bauthor{\bsnm{{Winkel}}, \binits{B.}},
\bauthor{\bsnm{{Bailin}}, \binits{J.}},
\bauthor{\bsnm{{Calabretta}}, \binits{M.R.}},
\bauthor{\bsnm{{Dedes}}, \binits{L.}},
\bauthor{\bsnm{{Ford}}, \binits{H.A.}},
\bauthor{\bsnm{{Gibson}}, \binits{B.K.}},
\bauthor{\bsnm{{Haud}}, \binits{U.}},
\bauthor{\bsnm{{Janowiecki}}, \binits{S.}},
\bauthor{\bsnm{{Kalberla}}, \binits{P.M.W.}},
\bauthor{\bsnm{{Lockman}}, \binits{F.J.}},
\bauthor{\bsnm{{McClure-Griffiths}}, \binits{N.M.}},
\bauthor{\bsnm{{Murphy}}, \binits{T.}},
\bauthor{\bsnm{{Nakanishi}}, \binits{H.}},
\bauthor{\bsnm{{Pisano}}, \binits{D.J.}},
\bauthor{\bsnm{{Staveley-Smith}}, \binits{L.}}:
\bjtitle{\aap}
\bvolume{594},
\bfpage{116}
(\byear{2016}).
\doiurl{https://doi.org/10.1051/0004-6361/201629178}.
\arxivurl{1610.06175}
\end{barticle}
\endbibitem

\bibitem[\protect\citeauthoryear{{Jansen} et~al.}{2001}]{2001A&A...365L...1J}
\begin{barticle}
\bauthor{\bsnm{{Jansen}}, \binits{F.}},
\bauthor{\bsnm{{Lumb}}, \binits{D.}},
\bauthor{\bsnm{{Altieri}}, \binits{B.}},
\bauthor{\bsnm{{Clavel}}, \binits{J.}},
\bauthor{\bsnm{{Ehle}}, \binits{M.}},
\bauthor{\bsnm{{Erd}}, \binits{C.}},
\bauthor{\bsnm{{Gabriel}}, \binits{C.}},
\bauthor{\bsnm{{Guainazzi}}, \binits{M.}},
\bauthor{\bsnm{{Gondoin}}, \binits{P.}},
\bauthor{\bsnm{{Much}}, \binits{R.}},
\bauthor{\bsnm{{Munoz}}, \binits{R.}},
\bauthor{\bsnm{{Santos}}, \binits{M.}},
\bauthor{\bsnm{{Schartel}}, \binits{N.}},
\bauthor{\bsnm{{Texier}}, \binits{D.}},
\bauthor{\bsnm{{Vacanti}}, \binits{G.}}:
\bjtitle{\aap}
\bvolume{365},
\bfpage{1}
(\byear{2001}).
\doiurl{https://doi.org/10.1051/0004-6361:20000036}
\end{barticle}
\endbibitem

\bibitem[\protect\citeauthoryear{{Jiang} et~al.}{2019}]{2019MNRAS.489.3436J}
\begin{barticle}
\bauthor{\bsnm{{Jiang}}, \binits{J.}},
\bauthor{\bsnm{{Fabian}}, \binits{A.C.}},
\bauthor{\bsnm{{Dauser}}, \binits{T.}},
\bauthor{\bsnm{{Gallo}}, \binits{L.}},
\bauthor{\bsnm{{Garc{\'\i}a}}, \binits{J.A.}},
\bauthor{\bsnm{{Kara}}, \binits{E.}},
\bauthor{\bsnm{{Parker}}, \binits{M.L.}},
\bauthor{\bsnm{{Tomsick}}, \binits{J.A.}},
\bauthor{\bsnm{{Walton}}, \binits{D.J.}},
\bauthor{\bsnm{{Reynolds}}, \binits{C.S.}}:
\bjtitle{\mnras}
\bvolume{489}(\bissue{3}),
\bfpage{3436}
(\byear{2019}).
\doiurl{https://doi.org/10.1093/mnras/stz2326}.
\arxivurl{1908.0727
2}
\end{barticle}
\endbibitem

\bibitem[\protect\citeauthoryear{{Jin} et~al.}{2012}]{2012MNRAS.425..907J}
\begin{barticle}
\bauthor{\bsnm{{Jin}}, \binits{C.}},
\bauthor{\bsnm{{Ward}}, \binits{M.}},
\bauthor{\bsnm{{Done}}, \binits{C.}}:
\bjtitle{\mnras}
\bvolume{425}(\bissue{2}),
\bfpage{907}
(\byear{2012}).
\doiurl{https://doi.org/10.1111/j.1365-2966.2012.21272.x}.
\arxivurl{1205.1846}
\end{barticle}
\endbibitem

\bibitem[\protect\citeauthoryear{{Krolik} et~al.}{1991}]{1991ApJ...371..541K}
\begin{barticle}
\bauthor{\bsnm{{Krolik}}, \binits{J.H.}},
\bauthor{\bsnm{{Horne}}, \binits{K.}},
\bauthor{\bsnm{{Kallman}}, \binits{T.R.}},
\bauthor{\bsnm{{Malkan}}, \binits{M.A.}},
\bauthor{\bsnm{{Edelson}}, \binits{R.A.}},
\bauthor{\bsnm{{Kriss}}, \binits{G.A.}}:
\bjtitle{\apj}
\bvolume{371},
\bfpage{541}
(\byear{1991}).
\doiurl{https://doi.org/10.1086/169918}
\end{barticle}
\endbibitem

\bibitem[\protect\citeauthoryear{{Magdziarz}
  et~al.}{1998}]{1998MNRAS.301..179M}
\begin{barticle}
\bauthor{\bsnm{{Magdziarz}}, \binits{P.}},
\bauthor{\bsnm{{Blaes}}, \binits{O.M.}},
\bauthor{\bsnm{{Zdziarski}}, \binits{A.A.}},
\bauthor{\bsnm{{Johnson}}, \binits{W.N.}},
\bauthor{\bsnm{{Smith}}, \binits{D.A.}}:
\bjtitle{\mnras}
\bvolume{301}(\bissue{1}),
\bfpage{179}
(\byear{1998}).
\doiurl{https://doi.org/10.1046/j.1365-8711.1998.02
015.x}
\end{barticle}
\endbibitem

\bibitem[\protect\citeauthoryear{{Makishima}
  et~al.}{1986}]{1986ApJ...308..635M}
\begin{barticle}
\bauthor{\bsnm{{Makishima}}, \binits{K.}},
\bauthor{\bsnm{{Maejima}}, \binits{Y.}},
\bauthor{\bsnm{{Mitsuda}}, \binits{K.}},
\bauthor{\bsnm{{Bradt}}, \binits{H.V.}},
\bauthor{\bsnm{{Remillard}}, \binits{R.A.}},
\bauthor{\bsnm{{Tuohy}}, \binits{I.R.}},
\bauthor{\bsnm{{Hoshi}}, \binits{R.}},
\bauthor{\bsnm{{Nakagawa}}, \binits{M.}}:
\bjtitle{\apj}
\bvolume{308},
\bfpage{635}
(\byear{1986}).
\doiurl{https://doi.org/10.1086/16
4534}
\end{barticle}
\endbibitem

\bibitem[\protect\citeauthoryear{{Mason} et~al.}{2001}]{2001A&A...365L..36M}
\begin{barticle}
\bauthor{\bsnm{{Mason}}, \binits{K.O.}},
\bauthor{\bsnm{{Breeveld}}, \binits{A.}},
\bauthor{\bsnm{{Much}}, \binits{R.}},
\bauthor{\bsnm{{Carter}}, \binits{M.}},
\bauthor{\bsnm{{Cordova}}, \binits{F.A.}},
\bauthor{\bsnm{{Cropper}}, \binits{M.S.}},
\bauthor{\bsnm{{Fordham}}, \binits{J.}},
\bauthor{\bsnm{{Huckle}}, \binits{H.}},
\bauthor{\bsnm{{Ho}}, \binits{C.}},
\bauthor{\bsnm{{Kawakami}}, \binits{H.}},
\bauthor{\bsnm{{Kennea}}, \binits{J.}},
\bauthor{\bsnm{{Kennedy}}, \binits{T.}},
\bauthor{\bsnm{{Mittaz}}, \binits{J.}},
\bauthor{\bsnm{{Pandel}}, \binits{D.}},
\bauthor{\bsnm{{Priedhorsky}}, \binits{W.C.}},
\bauthor{\bsnm{{Sasseen}}, \binits{T.}},
\bauthor{\bsnm{{Shirey}}, \binits{R.}},
\bauthor{\bsnm{{Smith}}, \binits{P.}},
\bauthor{\bsnm{{Vreux}}, \binits{J.-M.}}:
\bjtitle{\aap}
\bvolume{365},
\bfpage{36}
(\byear{2001}).
\doiurl{https://doi.org/10.1051/0004-6361:20000044}.
\arxivurl{astro-ph/0011216}
\end{barticle}
\endbibitem

\bibitem[\protect\citeauthoryear{{Matzeu} et~al.}{2020}]{2020MNRAS.497.2352M}
\begin{barticle}
\bauthor{\bsnm{{Matzeu}}, \binits{G.A.}},
\bauthor{\bsnm{{Nardini}}, \binits{E.}},
\bauthor{\bsnm{{Parker}}, \binits{M.L.}},
\bauthor{\bsnm{{Reeves}}, \binits{J.N.}},
\bauthor{\bsnm{{Braito}}, \binits{V.}},
\bauthor{\bsnm{{Porquet}}, \binits{D.}},
\bauthor{\bsnm{{Middei}}, \binits{R.}},
\bauthor{\bsnm{{Kammoun}}, \binits{E.}},
\bauthor{\bsnm{{Lusso}}, \binits{E.}},
\bauthor{\bsnm{{Alston}}, \binits{W.N.}},
\bauthor{\bsnm{{Giustini}}, \binits{M.}},
\bauthor{\bsnm{{Lobban}}, \binits{A.P.}},
\bauthor{\bsnm{{Joyce}}, \binits{A.M.}},
\bauthor{\bsnm{{Igo}}, \binits{Z.}},
\bauthor{\bsnm{{Ebrero}}, \binits{J.}},
\bauthor{\bsnm{{Ballo}}, \binits{L.}},
\bauthor{\bsnm{{Santos-Lle{\'o}}}, \binits{M.}},
\bauthor{\bsnm{{Schartel}}, \binits{N.}}:
\bjtitle{\mnras}
\bvolume{497}(\bissue{2}),
\bfpage{2352}
(\byear{2020}).
\doiurl{https://doi.org/10.1093/mnras/staa2076}.
\arxivurl{2007.065
75}
\end{barticle}
\endbibitem

\bibitem[\protect\citeauthoryear{{Mehdipour}
  et~al.}{2011}]{2011A&A...534A..39M}
\begin{barticle}
\bauthor{\bsnm{{Mehdipour}}, \binits{M.}},
\bauthor{\bsnm{{Branduardi-Raymont}}, \binits{G.}},
\bauthor{\bsnm{{Kaastra}}, \binits{J.S.}},
\bauthor{\bsnm{{Petrucci}}, \binits{P.O.}},
\bauthor{\bsnm{{Kriss}}, \binits{G.A.}},
\bauthor{\bsnm{{Ponti}}, \binits{G.}},
\bauthor{\bsnm{{Blustin}}, \binits{A.J.}},
\bauthor{\bsnm{{Paltani}}, \binits{S.}},
\bauthor{\bsnm{{Cappi}}, \binits{M.}},
\bauthor{\bsnm{{Detmers}}, \binits{R.G.}},
\bauthor{\bsnm{{Steenbrugge}}, \binits{K.C.}}:
\bjtitle{\aap}
\bvolume{534},
\bfpage{39}
(\byear{2011}).
\doiurl{https://doi.org/10.10
51/0004-6361/201116875}.
\arxivurl{1107.0659}
\end{barticle}
\endbibitem

\bibitem[\protect\citeauthoryear{{Miller} et~al.}{2006}]{2006A&A...453L..13M}
\begin{barticle}
\bauthor{\bsnm{{Miller}}, \binits{L.}},
\bauthor{\bsnm{{Turner}}, \binits{T.J.}},
\bauthor{\bsnm{{Reeves}}, \binits{J.N.}},
\bauthor{\bsnm{{George}}, \binits{I.M.}},
\bauthor{\bsnm{{Porquet}}, \binits{D.}},
\bauthor{\bsnm{{Nandra}}, \binits{K.}},
\bauthor{\bsnm{{Dovciak}}, \binits{M.}}:
\bjtitle{\aap}
\bvolume{453}(\bissue{1}),
\bfpage{13}
(\byear{2006}).
\doiurl{https://doi.org/10.1051/0004-6361:20065276}.
\arxivurl{astro-ph/0605130}
\end{barticle}
\endbibitem

\bibitem[\protect\citeauthoryear{{Miniutti} et~al.}{2009}]{2009MNRAS.394..443M}
\begin{barticle}
\bauthor{\bsnm{{Miniutti}}, \binits{G.}},
\bauthor{\bsnm{{Ponti}}, \binits{G.}},
\bauthor{\bsnm{{Greene}}, \binits{J.E.}},
\bauthor{\bsnm{{Ho}}, \binits{L.C.}},
\bauthor{\bsnm{{Fabian}}, \binits{A.C.}},
\bauthor{\bsnm{{Iwasawa}}, \binits{K.}}:
\bjtitle{\mnras}
\bvolume{394}(\bissue{1}),
\bfpage{443}
(\byear{2009}).
\doiurl{https://doi.org/10.1111/j.1365-2966.2008.14334.x}.
\arxivurl{0812.1652}
\end{barticle}
\endbibitem

\bibitem[\protect\citeauthoryear{{Mitsuda} et~al.}{1984}]{1984PASJ...36..741M}
\begin{barticle}
\bauthor{\bsnm{{Mitsuda}}, \binits{K.}},
\bauthor{\bsnm{{Inoue}}, \binits{H.}},
\bauthor{\bsnm{{Koyama}}, \binits{K.}},
\bauthor{\bsnm{{Makishima}}, \binits{K.}},
\bauthor{\bsnm{{Matsuoka}}, \binits{M.}},
\bauthor{\bsnm{{Ogawara}}, \binits{Y.}},
\bauthor{\bsnm{{Shibazaki}}, \binits{N.}},
\bauthor{\bsnm{{Suzuki}}, \binits{K.}},
\bauthor{\bsnm{{Tanaka}}, \binits{Y.}},
\bauthor{\bsnm{{Hirano}}, \binits{T.}}:
\bjtitle{\pasj}
\bvolume{36},
\bfpage{741}
(\byear{1984})
\end{barticle}
\endbibitem

\bibitem[\protect\citeauthoryear{{Novikov} and
  {Thorne}}{1973}]{1973blho.conf..343N}
\begin{bchapter}
\bauthor{\bsnm{{Novikov}}, \binits{I.D.}},
\bauthor{\bsnm{{Thorne}}, \binits{K.S.}}:
In: \bbtitle{Black Holes (Les Astres Occlus)},
p. \bfpage{343}
(\byear{1973})
\end{bchapter}
\endbibitem

\bibitem[\protect\citeauthoryear{{Pal} et~al.}{2016}]{2016MNRAS.457..875P}
\begin{barticle}
\bauthor{\bsnm{{Pal}}, \binits{M.}},
\bauthor{\bsnm{{Dewangan}}, \binits{G.C.}},
\bauthor{\bsnm{{Misra}}, \binits{R.}},
\bauthor{\bsnm{{Pawar}}, \binits{P.K.}}:
\bjtitle{\mnras}
\bvolume{457},
\bfpage{875}
(\byear{2016}).
\doiurl{https://doi.org/10.
1093/mnras/stw009}.
\arxivurl{1601.00503}
\end{barticle}
\endbibitem

\bibitem[\protect\citeauthoryear{{Petrucci} et~al.}{2018}]{2018A&A...611A..59P}
\begin{barticle}
\bauthor{\bsnm{{Petrucci}}, \binits{P.-O.}},
\bauthor{\bsnm{{Ursini}}, \binits{F.}},
\bauthor{\bsnm{{De Rosa}}, \binits{A.}},
\bauthor{\bsnm{{Bianchi}}, \binits{S.}},
\bauthor{\bsnm{{Cappi}}, \binits{M.}},
\bauthor{\bsnm{{Matt}}, \binits{G.}},
\bauthor{\bsnm{{Dadina}}, \binits{M.}},
\bauthor{\bsnm{{Malzac}}, \binits{J.}}:
\bjtitle{\aap}
\bvolume{611},
\bfpage{59}
(\byear{2018}).
\doiurl{https://doi.org/10.1051/0
004-6361/201731580}.
\arxivurl{1710.04940}
\end{barticle}
\endbibitem

\bibitem[\protect\citeauthoryear{{Piconcelli}
  et~al.}{2005}]{2005A&A...432...15P}
\begin{barticle}
\bauthor{\bsnm{{Piconcelli}}, \binits{E.}},
\bauthor{\bsnm{{Jimenez-Bail{\'o}n}}, \binits{E.}},
\bauthor{\bsnm{{Guainazzi}}, \binits{M.}},
\bauthor{\bsnm{{Schartel}}, \binits{N.}},
\bauthor{\bsnm{{Rodr{\'\i}guez-Pascual}}, \binits{P.M.}},
\bauthor{\bsnm{{Santos-Lle{\'o}}}, \binits{M.}}:
\bjtitle{\aap}
\bvolume{432}(\bissue{1}),
\bfpage{15}
(\byear{2005}).
\doiurl{https://doi.org/10.1051
/0004-6361:20041621}.
\arxivurl{astro-ph/0411051}
\end{barticle}
\endbibitem

\bibitem[\protect\citeauthoryear{{Pounds} et~al.}{1986}]{1986MNRAS.218..685P}
\begin{barticle}
\bauthor{\bsnm{{Pounds}}, \binits{K.A.}},
\bauthor{\bsnm{{Warwick}}, \binits{R.S.}},
\bauthor{\bsnm{{Culhane}}, \binits{J.L.}},
\bauthor{\bsnm{{de Korte}}, \binits{P.A.J.}}:
\bjtitle{\mnras}
\bvolume{218},
\bfpage{685}
(\byear{1986}).
\doiurl{https://doi.org/10.1093/mnras/218.4.685}
\end{barticle}
\endbibitem

\bibitem[\protect\citeauthoryear{{Reeves} et~al.}{2008}]{2008MNRAS.385L.108R}
\begin{barticle}
\bauthor{\bsnm{{Reeves}}, \binits{J.}},
\bauthor{\bsnm{{Done}}, \binits{C.}},
\bauthor{\bsnm{{Pounds}}, \binits{K.}},
\bauthor{\bsnm{{Terashima}}, \binits{Y.}},
\bauthor{\bsnm{{Hayashida}}, \binits{K.}},
\bauthor{\bsnm{{Anabuki}}, \binits{N.}},
\bauthor{\bsnm{{Uchino}}, \binits{M.}},
\bauthor{\bsnm{{Turner}}, \binits{M.}}:
\bjtitle{\mnras}
\bvolume{385}(\bissue{1}),
\bfpage{108}
(\byear{2008}).
\doiurl{https://doi.org/10.1111
/j.1745-3933.2008.00443.x}.
\arxivurl{0801.1587}
\end{barticle}
\endbibitem

\bibitem[\protect\citeauthoryear{{Robinson} et~al.}{2019}]{2019ApJ...880...68R}
\begin{barticle}
\bauthor{\bsnm{{Robinson}}, \binits{J.H.}},
\bauthor{\bsnm{{Bentz}}, \binits{M.C.}},
\bauthor{\bsnm{{Johnson}}, \binits{M.C.}},
\bauthor{\bsnm{{Courtois}}, \binits{H.M.}},
\bauthor{\bsnm{{Ou-Yang}}, \binits{B.}}:
\bjtitle{\apj}
\bvolume{880}(\bissue{2}),
\bfpage{68}
(\byear{2019}).
\doiurl{https://doi.org/10.3847/1538-4357/ab29f9}.
\arxivurl{1906.07062}
\end{barticle}
\endbibitem

\bibitem[\protect\citeauthoryear{{Schurch} and
  {Done}}{2007}]{2007MNRAS.381.1413S}
\begin{barticle}
\bauthor{\bsnm{{Schurch}}, \binits{N.J.}},
\bauthor{\bsnm{{Done}}, \binits{C.}}:
\bjtitle{\mnras}
\bvolume{381}(\bissue{4}),
\bfpage{1413}
(\byear{2007}).
\doiurl{https://doi.org/10.1111/j.1365-2966.2007.12
336.x}.
\arxivurl{0706.1885}
\end{barticle}
\endbibitem

\bibitem[\protect\citeauthoryear{{Shakura} and
  {Sunyaev}}{1973}]{1973A&A....24..337S}
\begin{barticle}
\bauthor{\bsnm{{Shakura}}, \binits{N.I.}},
\bauthor{\bsnm{{Sunyaev}}, \binits{R.A.}}:
\bjtitle{\aap}
\bvolume{500},
\bfpage{33}
(\byear{1973})
\end{barticle}
\endbibitem

\bibitem[\protect\citeauthoryear{{Singh} et~al.}{1985}]{1985ApJ...297..633S}
\begin{barticle}
\bauthor{\bsnm{{Singh}}, \binits{K.P.}},
\bauthor{\bsnm{{Garmire}}, \binits{G.P.}},
\bauthor{\bsnm{{Nousek}}, \binits{J.}}:
\bjtitle{\apj}
\bvolume{297},
\bfpage{633}
(\byear{1985}).
\doiurl{https://doi.org/10.1086/163560}
\end{barticle}
\endbibitem

\bibitem[\protect\citeauthoryear{{Str{\"u}der}
  et~al.}{2001}]{2001A&A...365L..18S}
\begin{barticle}
\bauthor{\bsnm{{Str{\"u}der}}, \binits{L.}},
\bauthor{\bsnm{{Briel}}, \binits{U.}},
\bauthor{\bsnm{{Dennerl}}, \binits{K.}},
\bauthor{\bsnm{{Hartmann}}, \binits{R.}},
\bauthor{\bsnm{{Kendzio-
rra}}, \binits{E.}},
\bauthor{\bsnm{{Meidinger}}, \binits{N.}},
\bauthor{\bsnm{{Pfeffermann}}, \binits{E.}},
\bauthor{\bsnm{{Reppin}}, \binits{C.}},
\bauthor{\bsnm{{Aschenbach}}, \binits{B.}},
\bauthor{\bsnm{{Bornemann}}, \binits{W.}},
\bauthor{\bsnm{{Br{\"a}uninger}}, \binits{H.}},
\bauthor{\bsnm{{Burkert}}, \binits{W.}},
\bauthor{\bsnm{{Elender}}, \binits{M.}},
\bauthor{\bsnm{{Freyberg}}, \binits{M.}},
\bauthor{\bsnm{{Haberl}}, \binits{F.}},
\bauthor{\bsnm{{Hartner}}, \binits{G.}},
\bauthor{\bsnm{{Heuschmann}}, \binits{F.}},
\bauthor{\bsnm{{Hippmann}}, \binits{H.}},
\bauthor{\bsnm{{Kastelic}}, \binits{E.}},
\bauthor{\bsnm{{Kemmer}}, \binits{S.}},
\bauthor{\bsnm{{Kettenring}}, \binits{G.}},
\bauthor{\bsnm{{Kink}}, \binits{W.}},
\bauthor{\bsnm{{Krause}}, \binits{N.}},
\bauthor{\bsnm{{M{\"u}ller}}, \binits{S.}},
\bauthor{\bsnm{{Oppitz}}, \binits{A.}},
\bauthor{\bsnm{{Pietsch}}, \binits{W.}},
\bauthor{\bsnm{{Popp}}, \binits{M.}},
\bauthor{\bsnm{{Predehl}}, \binits{P.}},
\bauthor{\bsnm{{Read}}, \binits{A.}},
\bauthor{\bsnm{{Stephan}}, \binits{K.H.}},
\bauthor{\bsnm{{St{\"o}tter}}, \binits{D.}},
\bauthor{\bsnm{{Tr{\"u}mper}}, \binits{J.}},
\bauthor{\bsnm{{Holl}}, \binits{P.}},
\bauthor{\bsnm{{Kemmer}}, \binits{J.}},
\bauthor{\bsnm{{Soltau}}, \binits{H.}},
\bauthor{\bsnm{{St{\"o}tter}}, \binits{R.}},
\bauthor{\bsnm{{Weber}}, \binits{U.}},
\bauthor{\bsnm{{Weichert}}, \binits{U.}},
\bauthor{\bsnm{{von Zanthier}}, \binits{C.}},
\bauthor{\bsnm{{Carathanassis}}, \binits{D.}},
\bauthor{\bsnm{{Lutz}}, \binits{G.}},
\bauthor{\bsnm{{Richter}}, \binits{R.H.}},
\bauthor{\bsnm{{Solc}}, \binits{P.}},
\bauthor{\bsnm{{B{\"o}ttcher}}, \binits{H.}},
\bauthor{\bsnm{{Kuster}}, \binits{M.}},
\bauthor{\bsnm{{Staubert}}, \binits{R.}},
\bauthor{\bsnm{{Abbey}}, \binits{A.}},
\bauthor{\bsnm{{Holland}}, \binits{A.}},
\bauthor{\bsnm{{Turner}}, \binits{M.}},
\bauthor{\bsnm{{Balasini}}, \binits{M.}},
\bauthor{\bsnm{{Bignami}}, \binits{G.F.}},
\bauthor{\bsnm{{La Palombara}}, \binits{N.}},
\bauthor{\bsnm{{Villa}}, \binits{G.}},
\bauthor{\bsnm{{Buttler}}, \binits{W.}},
\bauthor{\bsnm{{Gianini}}, \binits{F.}},
\bauthor{\bsnm{{Lain{\'e}}}, \binits{R.}},
\bauthor{\bsnm{{Lumb}}, \binits{D.}},
\bauthor{\bsnm{{Dhez}}, \binits{P.}}:
\bjtitle{\aap}
\bvolume{365},
\bfpage{18}
(\byear{2001}).
\doiurl{https://doi.org/10.1051/0004-6361:20000066}
\end{barticle}
\endbibitem

\bibitem[\protect\citeauthoryear{{Tanaka} et~al.}{1995}]{1995Natur.375..659T}
\begin{barticle}
\bauthor{\bsnm{{Tanaka}}, \binits{Y.}},
\bauthor{\bsnm{{Nandra}}, \binits{K.}},
\bauthor{\bsnm{{Fabian}}, \binits{A.C.}},
\bauthor{\bsnm{{Inoue}}, \binits{H.}},
\bauthor{\bsnm{{Otani}}, \binits{C.}},
\bauthor{\bsnm{{Dotani}}, \binits{T.}},
\bauthor{\bsnm{{Hayashida}}, \binits{K.}},
\bauthor{\bsnm{{Iwasawa}}, \binits{K.}},
\bauthor{\bsnm{{Kii}}, \binits{T.}},
\bauthor{\bsnm{{Kunieda}}, \binits{H.}},
\bauthor{\bsnm{{Makino}}, \binits{F.}},
\bauthor{\bsnm{{Matsuoka}}, \binits{M.}}:
\bjtitle{\nat}
\bvolume{375}(\bissue{6533}),
\bfpage{659}
(\byear{1995}).
\doiurl{https://doi.org/10.1038/375659a0}
\end{barticle}
\endbibitem

\bibitem[\protect\citeauthoryear{{Tripathi} et~al.}{2019}]{2019MNRAS.488.4831T}
\begin{barticle}
\bauthor{\bsnm{{Tripathi}}, \binits{S.}},
\bauthor{\bsnm{{Waddell}}, \binits{S.G.H.}},
\bauthor{\bsnm{{Gallo}}, \binits{L.C.}},
\bauthor{\bsnm{{Welsh}}, \binits{W.F.}},
\bauthor{\bsnm{{Chiang}}, \binits{C.-Y.}}:
\bjtitle{\mnras}
\bvolume{488}(\bissue{4}),
\bfpage{4831}
(\byear{2019}).
\doiurl{https://doi.org/10.1093/mnras/stz1988}.

\arxivurl{1907.07048}
\end{barticle}
\endbibitem

\bibitem[\protect\citeauthoryear{{Turner} et~al.}{2001}]{2001A&A...365L..27T}
\begin{barticle}
\bauthor{\bsnm{{Turner}}, \binits{M.J.L.}},
\bauthor{\bsnm{{Abbey}}, \binits{A.}},
\bauthor{\bsnm{{Arnaud}}, \binits{M.}},
\bauthor{\bsnm{{Balasini}}, \binits{M.}},
\bauthor{\bsnm{{Barbera}}, \binits{M.}},
\bauthor{\bsnm{{Belsole}}, \binits{E.}},
\bauthor{\bsnm{{Bennie}}, \binits{P.J.}},
\bauthor{\bsnm{{Bernard}}, \binits{J.P.}},
\bauthor{\bsnm{{Bignami}}, \binits{G.F.}},
\bauthor{\bsnm{{Boer}}, \binits{M.}},
\bauthor{\bsnm{{Briel}}, \binits{U.}},
\bauthor{\bsnm{{Butler}}, \binits{I.}},
\bauthor{\bsnm{{Cara}}, \binits{C.}},
\bauthor{\bsnm{{Chabaud}}, \binits{C.}},
\bauthor{\bsnm{{Cole}}, \binits{R.}},
\bauthor{\bsnm{{Collura}}, \binits{A.}},
\bauthor{\bsnm{{Conte}}, \binits{M.}},
\bauthor{\bsnm{{Cros}}, \binits{A.}},
\bauthor{\bsnm{{Denby}}, \binits{M.}},
\bauthor{\bsnm{{Dhez}}, \binits{P.}},
\bauthor{\bsnm{{Di Coco}}, \binits{G.}},
\bauthor{\bsnm{{Dowson}}, \binits{J.}},
\bauthor{\bsnm{{Ferrando}}, \binits{P.}},
\bauthor{\bsnm{{Ghizzardi}}, \binits{S.}},
\bauthor{\bsnm{{Gianotti}}, \binits{F.}},
\bauthor{\bsnm{{Goodall}}, \binits{C.V.}},
\bauthor{\bsnm{{Gretton}}, \binits{L.}},
\bauthor{\bsnm{{Griffiths}}, \binits{R.G.}},
\bauthor{\bsnm{{Hainaut}}, \binits{O.}},
\bauthor{\bsnm{{Hochedez}}, \binits{J.F.}},
\bauthor{\bsnm{{Holland}}, \binits{A.D.}},
\bauthor{\bsnm{{Jourdain}}, \binits{E.}},
\bauthor{\bsnm{{Kendziorra}}, \binits{E.}},
\bauthor{\bsnm{{Lagostina}}, \binits{A.}},
\bauthor{\bsnm{{Laine}}, \binits{R.}},
\bauthor{\bsnm{{La Palombara}}, \binits{N.}},
\bauthor{\bsnm{{Lortholary}}, \binits{M.}},
\bauthor{\bsnm{{Lumb}}, \binits{D.}},
\bauthor{\bsnm{{Marty}}, \binits{P.}},
\bauthor{\bsnm{{Molendi}}, \binits{S.}},
\bauthor{\bsnm{{Pigot}}, \binits{C.}},
\bauthor{\bsnm{{Poindron}}, \binits{E.}},
\bauthor{\bsnm{{Pounds}}, \binits{K.A.}},
\bauthor{\bsnm{{Reeves}}, \binits{J.N.}},
\bauthor{\bsnm{{Reppin}}, \binits{C.}},
\bauthor{\bsnm{{Rothenflug}}, \binits{R.}},
\bauthor{\bsnm{{Salvetat}}, \binits{P.}},
\bauthor{\bsnm{{Sauvageot}}, \binits{J.L.}},
\bauthor{\bsnm{{Schmitt}}, \binits{D.}},
\bauthor{\bsnm{{Sembay}}, \binits{S.}},
\bauthor{\bsnm{{Short}}, \binits{A.D.T.}},
\bauthor{\bsnm{{Spragg}}, \binits{J.}},
\bauthor{\bsnm{{Stephen}}, \binits{J.}},
\bauthor{\bsnm{{Str{\"u}der}}, \binits{L.}},
\bauthor{\bsnm{{Tiengo}}, \binits{A.}},
\bauthor{\bsnm{{Trifoglio}}, \binits{M.}},
\bauthor{\bsnm{{Tr{\"u}mper}}, \binits{J.}},
\bauthor{\bsnm{{Vercellone}}, \binits{S.}},
\bauthor{\bsnm{{Vigroux}}, \binits{L.}},
\bauthor{\bsnm{{Villa}}, \binits{G.}},
\bauthor{\bsnm{{Ward}}, \binits{M.J.}},
\bauthor{\bsnm{{Whitehead}}, \binits{S.}},
\bauthor{\bsnm{{Zonca}}, \binits{E.}}:
\bjtitle{\aap}
\bvolume{365},
\bfpage{27}
(\byear{2001}).
\doiurl{https://doi.org/10.1051/0004-6361:20000087}.
\arxivurl{astro-ph/0011498}
\end{barticle}
\endbibitem

\bibitem[\protect\citeauthoryear{{Walter} and
  {Fink}}{1993}]{1993A&A...274..105W}
\begin{barticle}
\bauthor{\bsnm{{Walter}}, \binits{R.}},
\bauthor{\bsnm{{Fink}}, \binits{H.H.}}:
\bjtitle{\aap}
\bvolume{274},
\bfpage{105}
(\byear{1993})
\end{barticle}
\endbibitem

\bibitem[\protect\citeauthoryear{{Walton} et~al.}{2013}]{2013MNRAS.428.2901W}
\begin{barticle}
\bauthor{\bsnm{{Walton}}, \binits{D.J.}},
\bauthor{\bsnm{{Nardini}}, \binits{E.}},
\bauthor{\bsnm{{Fabian}}, \binits{A.C.}},
\bauthor{\bsnm{{Gallo}}, \binits{L.C.}},
\bauthor{\bsnm{{Reis}}, \binits{R.C.}}:
\bjtitle{\mnras}
\bvolume{428}(\bissue{4}),
\bfpage{2901}
(\byear{2013}).
\doiurl{https://doi.org/10.1093/mnras/sts227}.
\arxivurl{1210.4593}
\end{barticle}
\endbibitem

\bibitem[\protect\citeauthoryear{{Wilms} et~al.}{2000}]{2000ApJ...542..914W}
\begin{barticle}
\bauthor{\bsnm{{Wilms}}, \binits{J.}},
\bauthor{\bsnm{{Allen}}, \binits{A.}},
\bauthor{\bsnm{{McCray}}, \binits{R.}}:
\bjtitle{\apj}
\bvolume{542}(\bissue{2}),
\bfpage{914}
(\byear{2000}).
\doiurl{https://doi.org/10.1086/317016}.
\arxivurl{astro-ph/0008425}
\end{barticle}
\endbibitem

\bibitem[\protect\citeauthoryear{{Zdziarski}
  et~al.}{1996}]{1996MNRAS.283..193Z}
\begin{barticle}
\bauthor{\bsnm{{Zdziarski}}, \binits{A.A.}},
\bauthor{\bsnm{{Johnson}}, \binits{W.N.}},
\bauthor{\bsnm{{Magdziarz}}, \binits{P.}}:
\bjtitle{\mnras}
\bvolume{283}(\bissue{1}),
\bfpage{193}
(\byear{1996}).
\doiurl{https://doi.org/10.10
93/mnras/283.1.193}.
\arxivurl{astro-ph/9607015}
\end{barticle}
\endbibitem

\bibitem[\protect\citeauthoryear{{{\.Z}ycki}
  et~al.}{1999}]{1999MNRAS.309..561Z}
\begin{barticle}
\bauthor{\bsnm{{{\.Z}ycki}}, \binits{P.T.}},
\bauthor{\bsnm{{Done}}, \binits{C.}},
\bauthor{\bsnm{{Smith}}, \binits{D.A.}}:
\bjtitle{\mnras}
\bvolume{309}(\bissue{3}),
\bfpage{561}
(\byear{1999}).
\doiurl{https://doi.org/10.1046/j.1365-8711.1999.02885.x}.
\arxivurl{astro-ph/9904304}
\end{barticle}
\endbibitem

\end{thebibliography}

\end{document}